\documentclass[12pt,preprint]{aastex}

\def\etal{{et~al}\hbox{.}}

\slugcomment{Submitted to ApJ: April 12, 2003; Version: \today}

\shorttitle{Combining WMAP and SDSS Quasar Data on Reionization}
\shortauthors{Chiu, Fan, \& Ostriker}

\begin{document}


\title{
Combining WMAP and SDSS Quasar Data on Reionization\\
Constrains Cosmological Parameters and\\ 
the Star Formation Efficiency
}


\author{Weihsueh A. Chiu}
\affil{U.S. Environmental Protection Agency, 
Washington, DC  20460}
\email{chiu@astro.princeton.edu}

\author{Xiaohui Fan}
\affil{Steward Observatory, The University of Arizona, Tucson, AZ 85721}
\email{fan@as.arizona.edu}
\and

\author{Jeremiah P. Ostriker}
\affil{Princeton University Observatory, Princeton, NJ 08544}
\email{jpo@astro.princeton.edu}




\begin{abstract}
We present constraints on cosmological and star formation parameters 
based on combining observations of the Wilkinson Microwave Anisotropy
Probe (WMAP) and high-redshift quasars from the Sloan Digital 
Sky Survey (SDSS).  
We use a semi-analytic model for reionization \citep{Chiu00}
that takes into account 
a number of important physical processes both within collapsing halos 
(e.g., H$_2$ cooling) and in the intergalactic medium (e.g., 
H$_2$ cooling, Compton cooling, and photoionization heating).  
We find that the Gunn-Peterson absorption
data provide tight constraints on the power spectrum at small
scales in a manner analogous to that derived from the cluster mass function.
Assuming that the efficiency of producing UV photons per baryon is 
constant, the constraint takes on the form 
$\sigma_8\Omega_0^{0.5} \approx 0.33$  
in a flat, $\Lambda$-dominated universe with
$h=0.72$, $n=0.99$, and $\Omega_b h^2 = 0.024$. However, the
calculated optical depth to electron scattering of $\tau_{\rm es} \sim 0.06$ 
is well below the value found by WMAP of $0.17\pm (0.04 \sim 0.07)$.
Since the WMAP constraints on $\tau_{\rm es}$ are somewhat degenerate
with the value of the spectral index $n$ \citep{Setal03},
we then permit the primordial spectral index $n$ to float and
fixing the best fit WMAP determination of $\Omega_0 h^2 = 0.14$,
while normalizing the power spectrum using WMAP.
In addition, we allow the UV-efficiency to have time-dependence.
Combining the WMAP constraints with the quasar transmission data,
our analysis then favors a model with $\tau_{\rm es}=0.11^{+0.02}_{-0.03}$,
$n = 0.96^{+0.02}_{-0.03}$, implying $\sigma_8=0.83^{+0.03}_{-0.05}$
(all at 95\% confidence), 
and an effective UV-efficiency 
that was at least $\sim 10\times$ greater at $z \gg 6$.  
The implied UV-efficiencies is not 
unreasonable for stars, spanning the range 
from $10^{-5.5} \sim 10^{-4}$.
These results indicate that the quasar and WMAP observations 
are consistent.  
If future observations confirm an optical depth
to electron scattering $\tau_{\rm es} \sim 0.1$, then it 
would appear that no more ``exotic'' sources of UV-photons, 
such as mini-quasars or AGNs,  are 
necessary; but our analysis indicates that a determination
of $\tau_{\rm es} \gtrsim 0.17$ would require a more radical
solution.  
\end{abstract}




\section{Introduction}

The primieval spectrum of cosmic density fluctuations on large scales 
is determinable with great precision from cosmic background
radiation measurements \citep{Page03}.  In order to determine the total
relevant spectrum, information on small scales is also necessary;
these become nonlinear at early times, so information
concerning reionization in the epochs $20 \lesssim z \lesssim 6$
provide vital clues.

Recent observations of high-redshift quasars have provided the 
first observational signatures of the epoch of reionization.  
Spectra of quasars at redshift $z \lesssim 6$ indicate that
the universe was almost fully ionized up to $z \sim 6$, since 
even a small neutral fraction in the intergalactic medium (IGM)
would have led to complete absorption of a quasar's 
continuum radiation.  However, the first absorption spectra of quasars 
at higher redshift indicate that the abundance of neutral 
hydrogen increases significantly for $z \gtrsim 6$.  

Reionization has generally been assumed to be caused by ionizing photons
created in early generations of stars and/or quasars.  Given this
premise, hydrodynamic simulations as well as semi-analytic 
calculations seem to indicate that the process of reionization 
should occur several distinct stages.  First, cosmological gas 
falls into deep enough potential wells (caused by dark matter halos), 
so that they can cool and
collapse to high enough densities to produce stars and/or quasars.
The UV photons produced in this process ionizes the local 
surroundings, first within the halo itself and then
outside the halo creating cosmological H~II regions.  This
is generally referred to as the ``pre-overlap'' stage.  As the 
abundance of these H~II regions increases (due to additional 
``galaxy'' formation), they eventually start to ``overlap'' so that
gas in the IGM becomes exposed to multiple sources of ionizing 
radiation.  After this ``overlap'' stage, the IGM becomes optically
thin except for inside self-shielded, high-density clouds
(those without ionizing sources).  

In this picture, the reionization history 
of the universe depends on both the growth of density perturbations
and the efficiency of star/quasar formation.  The former is a complex
function of the standard cosmological parameters --- the density
of the universe $\Omega_0$, the baryon abundance $\Omega_b$, 
the expansion rate $H_0$, and the mass power spectral index $n$
and normalization $\sigma_8$.  The latter can be calculated with
some precision based on atomic physics (e.g., cooling rates), 
but ultimately depends on some unknown parameters relating to 
the efficiency of turning mass into UV photons that can escape 
into the IGM.  

To this overall picture has been added the recent 
observations by the Wilkinson Microwave Anisotropy Probe (WMAP)
satellite \citep{WMAP}.  In particular, the high values 
of the electron optical depth to last scattering 
$\tau_{\rm es} = 0.17 \pm 0.04$ \citep{Kogut03}
and $\tau_{\rm es} = 0.17 \pm 0.07$ \citep{Setal03}
seem to indicate a much earlier epoch or reionization of 
$z_{\rm rei} = 17\pm 5$ \citep{Setal03}.  
How can these measurements be reconciled with the 
quasar Gunn-Peterson observations?  Do they indicate a source
of ionizing photons that cannot be accounted for through
standard star formation?

In this paper, we first examine how quasar transmission
measurements constrain cosmological parameters.  In particular,
we use the fact that the first generation of UV-generating objects
are in fact the tail of the distribution --- the rare events that
have collapsed to high enough density to produce stars and/or quasars.
They thus provide a unique probe of the small scale power spectrum 
(at $\lesssim 1$~Mpc scales) in an analogous way that the X-ray 
cluster mass function probes $\sim 10$~Mpc scales.  Our constraints
are derived from quasars absorption measurements, using theoretical
predictions based on the detailed semi-analytic model developed by 
\citet{Chiu00}, with some improvements derived from recent 
hydrodynamic and semi-analytical work on reionization.  

Second, we address the question of the consistency of the WMAP
and quasar measurements, paying particular attention to the
degeneracy in the WMAP data between the optical depth $\tau_{\rm es}$
and the spectral index $n$. Combining the WMAP constraints 
with the quasar transmission data,
our analysis favors a model with somewhat lower 
$\tau_{\rm es}$ and $n$ than implied by the WMAP data 
alone, but that are still consistent at the 1-$\sigma$ level.

The organization of this paper is as follows.
In \S~2, we describe the modeling of reionization 
and Gunn-Peterson absorption.  In \S~3, 
we compare observations from both quasars and WMAP to 
model predictions in order to constrain
cosmological parameters.  In \S~4, we discuss 
our results and discuss their implications.  We summarize and
conclude in \S~5.  

\section{Modeling Reionization and Gunn-Peterson Absorption}

\subsection{Summary of the Model}

The details of the semianalytic model are described in
\citet{Chiu00}; the basic principles are summarized here.  
It is based on a two-phase model of the universe in 
which a statistical filling factor for ionized gas 
is self-consistently calculated. 
It is assumed that the 
cold, neutral phase has no sources, and evolves passively
with the expansion of the universe.  The hot, ionized phase
contains the ionizing sources and evolves in line with 
local particle and energy conservation averaged over the phase.  
The temperature of each phase is calculated using standard
physics, with photoheating as the source of heat in the ionized
phase and cooling via multiple mechanisms, including H$_2$, 
atomic lines, and Compton scattering.  
The time evolution is determined by particle and
energy conservation.  
In each case, we only consider the regions outside of 
collapsed gas halos, which, of course will rise to the virial
temperature.  Such collapsed halos are calculated separately,
and considered potential sources of ionizing radiation.

The abundance and properties of these potential ionizing sources 
are calculated on the basis of 
the Press-Schechter formalism \citep{PS74}, 
constrained by the Jeans' criterion
(which utilizes the calculated gas temperatures) and by a 
cooling criterion (the cooling time must be less than the 
dynamical time).  Cooling in the halos includes the important
contributions from H$_2$ cooling.  
Halos that statisfy the Jeans criterion 
and that can cool efficiently are sources of ionizing radiation.
We calculate the luminosity of each ionizing
source using the Schmidt law
\begin{equation}
	L(M_b) = (1-f_\ast)M_b c^2 
	\epsilon_{\rm esc} 
	\epsilon_{\ast} \epsilon_{\rm UV} t_{\rm dyn}^{-1} ,
\end{equation}
where $f_\ast$ is the fraction of baryons already turned into stars
(a small correction), $M_b$ is the baryonic mass of a halo,
$\epsilon_{\rm esc}$ is the escape fraction from the halo, 
$\epsilon_{\ast}$ is a resolution factor (related to the 
fraction of gas that can form stars) determined through
calibration (see \citet{Chiu00} equation (28), and below), 
$\epsilon_{\rm UV}$
is the mass to UV efficiency (where we have absorbed the notation
in \citet{Chiu00} of $\epsilon_{\rm hm}\epsilon_{\rm UV}$ into 
a single efficiency), and $t_{\rm dyn}$ is the dynamical time of
the halo.  Note that this halo luminosity is somewhat simpler
than that used in \citet{Chiu00}, but is consistent with that used
in hydrodynamic simulations.  

Calibration was done by comparing 
the redshift of overlap, the ionizing intensity, and the 
fraction of baryonic mass in stars between 
the semi-analytic results and
the hydrodynamic simulation of Gnedin (2000ab).  This simulation 
was chosen because it is the only published simulation which 
simulates a statistically ``average'' universe (as opposed to 
an individual halo collapse) with sufficient resolution 
to follow the overlap process, and continues at least 
redshift 4.  The cosmological parameters were 
$\Omega_0 = 0.3$, $\Omega_\Lambda = 0.7$, $h=0.7$, $\Omega_b = 0.04$, 
$n = 1$, and $\sigma_8 = 0.9$.  
The data used for calibration from Gnedin (2000ab) are the redshift of
overlap $z_{\ast} \approx 7 $, the ionizing intensity of 
$J_{21} \approx 0.3$ at $z=4$, and the 
fraction of baryonic mass in stars 
$f_{\ast} \approx 0.04$ at $z=4$.  
The values of $\epsilon_{\ast}$ and $\epsilon_{\rm UV}$ 
in the semianalytic model were adjusted to best match 
these three values.
The results were  $\epsilon_{\ast} = 0.03$ and 
$\epsilon_{\rm UV} = 1.2 \times 10^{-5}$, which gave 
$z_{\ast} = 7$, $J_{21} = 0.6$, and $f_{\ast} = 0.05$.  
Note that Gnedin (2000ab) used $\epsilon_{\ast} = 0.05$ and
$\epsilon_{\rm UV} = 4 \times 10^{-5}$, 
which are remarkably similar to the values in the semianalytic model
given the differences in approach.  
In fact, because we take each halo as a whole, it should not be surprising
that $\epsilon_{\ast}$ is smaller for the semianalytic model
than for the hydrodynamic model.  

The determination of the escape fraction $\epsilon_{\rm esc}$ bears 
some additional discussion.  We use a simple Str\"omgren sphere
approximation (see Appendix) and derive
\begin{equation}
\epsilon_{\rm esc} \approx \left(1 - \eta\right)^2 ,
\end{equation}
where the quantity $\eta$ is given by
\begin{equation}
\eta = \mbox{Min}\left[1,\ \sqrt{\frac{\Delta_v \bar{\rho_b} \alpha_R}
	{3 \epsilon_{\ast} \epsilon_{\rm UV} 
	m_{\rm H}^2 c^2 E_0^{-1} t_{\rm dyn}^{-1}}}\right]
 ,
\end{equation}
where $\Delta_v \approx 178$ is the virial overdensity, $\bar{\rho_b}$
is the mean baryonic density, $\alpha_R$ is the recombination rate,
$m_{\rm H}$ is the hydrogen mass, and $E_0$ is 13.6 eV.  In this 
approximation, we assume the halo baryons have a $r^{-2}$ 
density run, the UV output per baryon is constant
in the halo, the halo is in ionization equilibrium, and 
recombinations determine the amount of photons absorbed locally. 

The ionizing source luminosity defined above, then, is the luminosity
seen by the IGM --- i.e., outside of the halos.  A further assumption
is made in the pre-overlap stage: that at any particular time, 
the ionized volume around an isolated source is linearly 
related to its UV luminosity.  The constant of proportionality is
determined through the global ionization and energy balance, but 
the size of the H~II regions as a function of luminosity is a linear
relation.  This assumption is consistent with previous 
work on cosmological H~II regions (e.g., Shapiro and Giroux 1987).

The outputs of the model that we use below are the filling factor 
$Q$, the temperature $T_4$, and the ionizing intensity $J_{21}$.

\subsection{Density Distribution of Cosmic Gas}

One of the most important determinants of how reionization 
evolves is the degree of gas clumping.  The clumping factor,
defined by 
\begin{equation} 
{\cal C} \equiv \langle \rho_b^2\rangle / \langle \rho_b\rangle^2 ,
\end{equation}
is important not only for determining the ionization balance
(and hence the filling factors), but also for 
determing the \emph{thermal} balance.  Note that this clumping
factor is only used in the ionized region.

In order to determine the clumping factor, the probability density
function (PDF) for the gas density must be known.  
\citet{MHR00} found that a good fit to the volume-weighted PDF 
as seen in hydrodynamical simulations is
\begin{equation}
P_V(\Delta)d\Delta = A \exp\left[
	-\frac{(\Delta^{-2/3} - C_0)^2}{2(\delta_0/3)^2}\right]
	\Delta^{-\beta} d\Delta
\end{equation}
where the overdensity $\Delta = \rho_b/\bar{\rho_b}$, 
$\delta_0 = 7.61(1+z)$ is related to the linear rms gas density 
fluctuation, and $\beta$ is related to the density
run at high densities.  The parameters $A$ and $C_0$ are found
by requiring the mass and volume to be normalized to unity.

This formula, with the parameterizations of $\beta$ and
$\delta_0$ found in \citet{MHR00}, has been used in many others in 
calculating characteristics of reionization
(e.g., Songaila and Cowie 2002, Fan \etal 2002,).  However, it is
not often noted that the parameterization is based on a single 
simulation with a particular set of cosmological parameters.  
In particular, \citet{MHR00} used the simulation reported in
\citet{MCOR96}, which was a $\Lambda$-CDM model with $\Omega_0=0.4$,
$\Omega_\Lambda = 0.6$, $\Omega_b h^2=0.015$, $h=0.65$, $n=1$, and 
$\sigma_8=0.79$.  Since in this paper we are considering variants in 
the background cosmology, it is certainly not sufficient to use
the \citet{MHR00} parameterization for the gas PDF.  In particular,
the amount of power at small scales will depend significantly on
all the cosmological parameters (especially $\sigma_8$).
We therefore analyzed a second simulation (which uses different
cosmological parameters) and to generate a second set of parameter
fits.  The second simulation is one by \citet{Cen02b} 
in which $\Omega_0=0.3$, $\Omega_\Lambda=0.7$, $\Omega_b h^2 = 0.017$,
$h = 0.67$, $n=1$, and $\sigma_8 = 0.9$.  The values of $\beta$ 
and $\delta_0$ are given for both \citet{MHR00} and \citet{Cen02b}
are provided in Table 1.  

As is clear from Table 1, the value of $\delta_0$ and to a lesser
extent $\beta$ depend on cosmology.  This is not surprising given
that $\delta_0$ \emph{must} depend on the power spectrum.  The 
cosmological dependence on $\beta$ is less clear, especially since
the best fit value of $\beta$ depends significantly on the density 
run in collapsed halos.  Because the \citet{Cen02b} simulation is at 
much higher resolution and includes more realistic physics, we 
simply fit $\beta$ to the Cen results as a function of redshift,
with a maximum of $\beta_{\rm max} = 2.5$, corresponding to an
isothermal sphere:
\begin{equation}
\beta \approx \mbox{Min}[2.5,\ 3.2 - 4.73/(1+z)],\qquad z \geq 4
\end{equation}
where we are only considering redshifts $z \geq 4$.

As for a prescription for finding $\delta_0$, we note that 
the gas PDF also predicts the fraction of mass in collapsed
virialized halos:
\begin{equation}
f_{\rm gas\ PDF}({\rm collapsed}) 
	\approx \int_{6\pi^2}^{\infty} \Delta P_V(\Delta)d\Delta .
\end{equation}
The integration point $6\pi^2$ 
is derived from the fact that 
for a singular isothermal sphere, the local overdensity density at 
the virial radius is $6\pi^2$. Therefore, we are approximately
taking into account all gas within virialized halos.
Similarly, from linear theory, we can use the Press-Schechter
formalism \citep{PS74} 
to find the same quantity if we know the correct filtering
radius $R_f$:
\begin{equation}
f_{\rm PS}({\rm collapsed}) \approx \sqrt{\frac{2}{\pi\sigma_{R_f}^2}} 
	\int_{\delta_c}^\infty \exp
	{\left[-\frac{\delta^2}{2\sigma_{R_f}}\right]} d\delta ,
\end{equation}
where $\delta_c \approx 1.69$ and $\sigma_{R_f}$ is the 
linear RMS mass fluctuation filtered with a tophat of radius $R_f$.
However, the mass fraction should be equal by the two calculations
\begin{equation}
f_{\rm gas\ PDF}({\rm collapsed}) = f_{\rm PS}({\rm collapsed}) .
\end{equation}
By analyzing both simulations, we find the following
relation leads to satisfactory results: 
\begin{equation}
R_f  \approx  R_J/4 ,
\end{equation}
where $R_J$ is the Jeans length
defined by
\begin{equation}
R_J = \sqrt{\frac{5 \pi k_B T}{12 G\bar{\rho}\mu m_H a^2}}
\end{equation}
and $T$ is the gas temperature, $\bar{\rho}$ is the mean total density,
$\mu$ is the baryons per particle. 
With this prescription for finding $\delta_0$ and $\beta$, we need only
use the normalization constraints to find $A$ and $C_0$ for an 
arbitrary background cosmology.  While by no means perfect, this 
procedure should account to first order the dependence of the gas PDF
on background cosmology, and is certainly more accurate than a 
treating the evolution of $\delta_0$ as independent of cosmology.

Given the gas PDF, one can determine the clumping factor by simple
integration over all densities.  However, in reality, there is a 
high density cut-off due to the fact that the very high density regions
are not participating in the ionization balance due to self-shielding.
However, the determinination of the cutoff in our prescription is
not difficult, since we have explicitly made a separation between
the ``in-halo'' and ``out-of-halo'' calculations.  
Thus, we calculate the clumping
factor by integrating
\begin{equation}
{\cal C} = \frac{\int_0^{6\pi^2} \Delta^2 P_V(\Delta)d\Delta}
	{\int_0^{6\pi^2} P_V(\Delta)d\Delta} .
\end{equation}
As above, we assume that gas with overdensity $> 6\pi^2$ is 
within collapsed halos.  

We should note that this treatment produces smaller clumping
factors ($\sim 5$) than typically used in semi-analytic treatments.
This is because we are essentially using a ``hybrid'' between
the reionization treatment of \citet{MHR00} and the usual
clumping factor methods.  Conceptually, \citet{MHR00} considered
a density-dependent ionization fraction --- they assumed that
the universe was ionized up to a critical density, above which
was neutral and self-shielded.  Here, we are assuming that
the collapsed halos are self-shielded, and then treat the
rest of the universe using the clumping factor approach.  As 
described in the appendix, at redshifts before full reionization,
halos are optically thick if they do not contain sources themselves.
By design, this treatment does not include screening and evaporation of
mini-halos (e.g., \citet{BL02}).

To summarize, the semi-analytic model actually has \emph{three}
phases: the ``interior of halos,'' the hot ionized H~II regions,
and the cold neutral regions.
The ``interior of halos'' phase includes all the 
gas in the universe that is in 
halos with baryonic mass greater than the Jeans mass.  
We assume that these regions have are virialized with isothermal sphere 
density runs, and therefore include all gas with overdensity
greater than $6\pi^2$.  
In this phase, the cooling criterion and the Schmidt law 
are used to calculate the ionizing source function, and 
the photoionizations and recombinations are
treated through the escape fraction calculation (a Str\"omgren
sphere approximation).  
Those photons that ``escape'' this phase then enter 
the ``second phase,'' which are the ionized H II regions.  
In this phase, a clumping
factor approach is used, with a high density cutoff at
$6\pi^2$.  The last phase is the cold neutral part of
the universe, which uses up ionizing photons only when ``converted''
into the ionized phase.

\subsection{Modeling Gunn-Peterson Absorption}

The statistics of Gunn-Peterson absorption has been reviewed by
numerous others (Miralda-Escud\'e,  Haehnelt,  and Rees, 2000;
McDonald and Miralda-Escud\'e 2001; 
Fan \etal 2002; Songaila and Cowie 2002).  
To summarize, we begin with the stardard 
optical depth at resonance for a gas with neutral
density $n_{\rm HI}$ at redshift $z$:
\begin{equation}
\tau = \frac{3 \Lambda_{\rm 2p \rightarrow 1s} \lambda_\alpha^3 
	n_{\rm HI}(z)}
	{8\pi H_0\sqrt{\Omega_0(1+z)^3 + \Omega_\Lambda}},
\end{equation}
where $\Lambda_{2p\rightarrow 1s}$ is the decay rate ($6.25\times 10^8$ 
sec$^{-1}$), $\lambda_\alpha$ is the Lyman-$\alpha$ wavelength 
($1.216 \times 10^{-5}$ cm), and the other symbols have their 
usual meaning in a cosmological context. 

Now, we make the standard assumptions of ionization equilibrium
and a uniform ionizing 
intensity in the ionized regions to
obtain the neutral fraction.  
Considering only photoionization and 
radiative recombination 
\begin{equation}
n_{\rm HI} \Gamma_{21} J_{21} = X(X+Y/4)/m_p^2\, R_4 T_{4,0}^{-0.7} 
	\bar{\rho}_b^2 \Delta^2 ,
\end{equation}
where $Y=0.24$ is the helium mass fraction 
(we assume helium is singly ionized),
$R_4 \approx 4.2\times 10^{-13} {\rm cm}^{3}\, {\rm s}^{-1}$ 
is the recombination rate at a temperature $T=10^4\, {\rm K}$, 
$T_{4,0} = T_0/10^4\, {\rm K}$ is the temperature at 
mean density in units of $10^4$ Kelvin, and 
$\Gamma_{21}$ is the photoionization
rate for $J_{21}=1$.
We use the method of \citet{HG97} to calculate
the temperature at mean density after reionization
(see also Hui and Haiman 2003).
Plugging in values for the various other constants 
($\Gamma_{21}=4.35\times 10^{-12}$~s$^{-1}$,
$m_p=1.67\times 10^{-24}$~g,
$\bar{\rho}_b = 1.88\times 10^{-29} (1+z)^3 \Omega_b h^2$~g~cm$^{-3}$)
gives 
\begin{equation}
n_{\rm HI} = 7.63 
	\times 10^{-12} 
	\frac{(\Omega_b h^2)^2 T_{4}^{-0.7} 
	\Delta^{2} (1+z)^6}{J_{21}} 
	{\rm cm}^{-3}
\end{equation}
or
\begin{equation}
\tau =
	\frac{0.316\,  
	\Omega_b^2\, h^3 (1+z)^6}
	{ \sqrt{\Omega_0 (1+z)^3 + \Omega_\Lambda}} 
	\frac{T_{4}^{-0.7}}{J_{21}}
	\Delta^{2}
\end{equation}
where $T_{4}$, $\Delta$,  
and $J_{21}$ depend on the redshift $z$, but only $\Delta$ also 
depends on spatial location.  
It is helpful to reformulate this for 
$\Omega_0(1+z)^3 \gg \Omega_\Lambda$, and scaled from 
$\Omega_b h^2 = 0.02$ and $\Omega_0 h^2 = 0.14$:
\begin{equation}
\tau =
	1.536 
	\left(\frac{1+z}{1+5.5}\right)^{4.5}
	\sqrt{\frac{0.14}{\Omega_0 h^2}}
	\left(\frac{\Omega_b\, h^2}{0.02}\right)^2
	\frac{T_{4}^{-0.7}}{J_{21}}
	\Delta^{2} .
\end{equation}
Defining the optical depth
for a uniform medium ($\Delta = 1$) $\tau_u$ is
\begin{equation}
\tau_u \equiv
	1.536 
	\left(\frac{1+z}{1+5.5}\right)^{4.5}
	\sqrt{\frac{0.14}{\Omega_0 h^2}}
	\left(\frac{\Omega_b\, h^2}{0.02}\right)^2
	\frac{T_{4}^{-0.7}}{J_{21}} ,
\end{equation}
we can define the mean transmitted flux ratio at a given
redshift
\begin{equation}
{\cal T}_z = \langle e^{-\tau_u \Delta^2} \rangle = 
	Q \int_0^{\infty} P_V(\Delta) 
	e^{-\tau_u \Delta^2}
	d\Delta,
\end{equation}
where $Q$ is the volume filling factor for the ionized regions
(we have now adopted the standard notation --- \citet{Chiu00} used
``$f$'' as the volume filling factor).  Note that strictly 
speaking, there should be a high density cutoff.  However,
since the exponential has a power of $\Delta^2$, 
for large values of $\Delta$, the optical depth is very large.   
The high-density end of the PDF contributes very 
little to the integral, and the results are thus 
not sensititive to this cutoff.  In fact, 
at the redshifts of interest, the integral is usually
dominated by values of $\Delta < 1$, as was noted by 
\citet{B01} and others.  

Our semi-analytic model provides the values of $Q$,  $T_4$, and $J_{21}$, 
and from the previous section, we have a prescription for 
determining $P_V(\Delta)$.  Thus, we have a complete model
from which to calculate the expected Gunn-Peterson absorption.

\section{Comparing Model Predictions and Observations}

Having calibrated our model using various hydrodynamic 
simulations, we are left with one 
free parameter, $\epsilon_{\rm UV}$, in addition to the 
cosmological parameters.  Below, 
we first consider the case where $\epsilon_{\rm UV}$ is 
constant and derive a constraint 
on $\sigma_8\Omega_0^{0.5}$, similar to those from rich 
clusters, and compare the results 
and predictions to those found by WMAP.  We consider 
joint constraints from WMAP and 
the SDSS quasars on the spectral index $n$ as well as 
the time-depedence of 
$\epsilon_{\rm UV}$.

\subsection{Combining SDSS Quasar Data}

The quasar data we use combine the compilation at 
$z_{\rm abs}\lesssim 5.5$
\citet{SC2002} (Table~\ref{tab:transmissiondatacowie}) 
and based on measurements of six SDSS quasars at $z>5.7$ 
from \citet{Becker01},  \citet{Fan03}, and 
\citet{White03} (Table~\ref{tab:transmissiondata}).  
Note that we did not 
include the J1044-0125 ($z=5.74$) data in Table~\ref{tab:transmissiondata} 
because it is
already incorporated into the compilation of \citet{SC2002}.

In order to combine these data, we must take into account
the fact that at redshifts
$z\lesssim 5.6$, the uncertainties in the transmission 
data are dominated by intrinsic scatter, while at higher 
redshifts, the uncertainties are dominated by measurement
errors.  Since our model predicts the \emph{mean} transmission,
our likelihood function must properly account for both
measurement error as well as the scatter.  
For a compilation of transmissions, such as that by 
\citet{SC2002}, the contribution of the 
data $D_c = \{(z_j, {\cal T}_j)\}$ to the likelihood is simply
\begin{eqnarray}
P({\cal T}_z|D_c) & = & \exp\left[-\frac{1}{2}\sum_j 
	\frac{({\cal T}_{z_j}-{\cal T}_j)^2}
	{\sigma_{j,\rm mean}^2}\right] ,\\
\sigma_{j,\rm mean}^2 & = & \sigma_{j,\rm scatter}^2 /\sqrt{N} .
\label{eq:likelihoodcompiled}
\end{eqnarray}
The contribution to the likelihood from the compiled data 
is estimated this way.

For the individual transmission data, especially in the case
where both measurement error and scatter are important, the simplest 
way to do this is to estimate scatter 
$\sigma_{\rm scatter}$ and
add in quadrature with the measurement error $\sigma_{\rm meas}$
to obtain the total uncertainty in the mean for each data point.  
In particular, for individual transmission data 
$D_i = \{(z_j,T_j)\}$with 
known (Gaussian) errors 
$\sigma_{\rm meas}$ and $\sigma_{\rm scatter}$, 
probability of a mean transmission as a function
of redshift ${\cal T}_z$ is
\begin{equation}
P({\cal T}_z|D_i) = \exp\left[-\frac{1}{2}\sum_j 
	\frac{({\cal T}_{z_j}- T_j)^2}
	{\sigma_{j,\rm meas}^2+\sigma_{j,\rm scatter}^2}\right] .
\label{eq:likelihoodindividual}
\end{equation}
Note that if the measurement error is negligible, then 
Equation \ref{eq:likelihoodindividual} 
reduces to Equation~(\ref{eq:likelihoodcompiled}).
To estimate $\sigma_{\rm scatter}$, we use the \citet{SC2002}
compilation for $z\lesssim 5.6$ (interpolated), and use the 
data themselves (binned) to determine $\sigma_{\rm scatter}$ 
at higher redshifts.  These results are also tabluated in 
Table~\ref{tab:transmissiondata}.  The total likelihood 
function, $\cal L$, then is given by the product of the two separate
likelihoods.

\subsection{SDSS Quasar Contraints with a Constant UV-Efficiency 
Compared with WMAP}

For this initial comparison, we 
fix the other cosmological parameters to their 
WMAP-only best fit values: $n=0.99$, 
$\Omega_b h^2=0.024$, and $h=0.72$.  We use the 
WMAP-only results from \citet{Setal03}, Table 1, to 
ensure clarity in the data underlying our analysis.  
Note, however, that these last two values have 
additional independent lines of support, as noted in
\citet{Setal03}.  

We use a Bayesian method to determine the cosmological 
constraints.  The priors we use are $\Omega_0 \in [0.15, 0.40]$ and
$\sigma_8 \in [0.5,1.0]$.  This region essentially bounds
the 99\% region reported for the combined CMB and cluster
analysis of \citet{MBBS02}.  

In order to constrain $\Omega_0$ and $\sigma_8$ separately from 
$\epsilon_{\rm UV}$, we marginalize over the latter parameter 
to yield the marginalized likelihood distribution:
\begin{equation}
{\cal L}(\Omega_0,\sigma_8) = \int 
	{\cal L}(\Omega_0,\sigma_8,\epsilon_{\rm UV}) d\epsilon_{\rm UV} .
\end{equation}
The central values and percentile limits are found by integrating
over ${\cal L}(\Omega_0,\sigma_8)$.  

The result of this analysis is presented in Figure \ref{fig:mainresult}, 
where we plot 
the likelihood contours in the plane $\Omega_0 - \sigma_8$.  
As can be seem by the figure,
there is a considerable degeneracy in this plane similar 
to that derived from rich clusters.  Thus, 
the cosmological constraint from Gunn-Peterson
absorption can be summarized by
\begin{equation}
\sigma_8 \Omega_0^{0.5} = 0.33 \pm 0.01 ,
\label{eq:mainresult}
\end{equation}
where the error term is statistical error only (68\%).  

Given
the complex physics of the reionization model, it is difficult
to determine a meaningful systematic error-bar.  However,
we did investigate the effects of changing $h$ and $n$.  
The net result was to shift the right-hand-side of equation 
(\ref{eq:mainresult}) so that
\begin{equation}
\sigma_8\Omega_0^{0.5} \approx (0.33\pm 0.01) 
	\left(\frac{0.72}{h}\right)^{0.9 + \frac{n-1}{2}} 
	\left(\frac{n}{0.99}\right)^{-0.85} ,
\end{equation}
so that lower levels of $h$ or $n$ increase the right-hand-side.
Thus, lowering $h$ or $n$ slightly would make the Gunn-Peterson 
constrain more consistent.

These results for a constant UV efficiency are 
somewhat discordant with WMAP's
marginalized value of $\sigma_8\Omega_0^{0.5} = 0.48 \pm 0.12$, 
although only at
the 1.25-$\sigma$ level.  They are  consistent 
with the cluster-determinations of \citet{Bahcall02}
of $\sigma_8\Omega_0^{0.6} = 0.33 \pm 0.03$ (note different exponent), 
although again at about the 1-$\sigma$ level.

The actual \emph{value} of $\epsilon_{\rm UV}$ bears some discussion.  
Although we have left this parameter as completely free, 
there certainly are astrophysical constraints on its value.  
In our 95\% region, we find that 
$\epsilon_{\rm UV} \sim 2 \times 10^{-5}$ (see left panel 
of Figure~\ref{fig:epsandtau}).  
For a Scalo mass function with metal-enriched stars (1/20th solar
metalicity), the value calculated from population synthesis 
is $\sim 5 \times 10^{-5}$ \citep{WL02}.  
Given the uncertainties and the relative simplicity of our model, 
the correspondence is quite remarkable.  

The more striking inconsistency is that in all cases, 
the optical depth to electron scattering,
\begin{equation}
\tau_{\rm es} \equiv \int_0^{1000} dz \frac{dt}{dz} c \sigma_{\rm T} n_e ,
\end{equation}
where $\sigma_{\rm T} = 6.652 \times 10^{-25} {\rm cm}^2$ 
is the Thomson cross-section and $n_e$ is the electron density,
is too low.  
Our model finds that $\tau_{\rm es} \sim 0.06$ (see right  panel 
of Figure~\ref{fig:epsandtau}), whereas \citet{Setal03} 
and \citet{Kogut03} report WMAP-only best fit values of 
$\tau_{\rm es} = 0.17\pm 0.07$ and $\tau_{\rm es} = 0.17\pm 0.04$, 
respectively. Several articles written in the wake of 
these results  (e.g., Haiman and Holder 2003; 
Ciardi, Ferrara, and White 2003; Cen 2003) 
have discussed this inconsistency and 
indicated that star formation must begun much earlier than 
previously thought.  We consider this issue in the following analysis.

\subsection{Combined WMAP and SDSS Constraints 
with a Time-Varying UV-Efficiency}

There are good astrophysical reasons to believe 
that $\epsilon_{\rm UV}$ may effectively 
increase with redshift.  For instance, the first generation of
stars would have been metal-free, and stellar models predict them
to have an significantly higher UV output per baryon (about $4\times$ 
higher was reported by \citet{WL02}).  In addition,
the creation of metals and henceforth dust, which would obscure 
ionizing sources, may also lead to lower \emph{effective} efficiencies
at lower redshift than at higher redshift.  

Thus we now consider constraints on $n$ and the time-dependence 
of $\epsilon_{\rm UV}$
from WMAP and SDSS quasars jointly.  To limit the dimensionality 
of the parameter space, we keep the baryon 
abundance and the Hubble constant at their WMAP best fit 
values, and in addition fix $\Omega_0 h^2 = 0.14$ to its 
WMAP best fit value.  For each value 
of $n$ and $\tau_{\rm es}$, WMAP predicts a unique best-fit
value of the normalization $\sigma_8$ (based on the code
provided by Verde \etal\ 2003 and the 
accompanying data files from
Hinshaw \etal\ 2003 and Kogut \etal\ 2003).  The normalization
given approximately by
(for our fixed values of $h$, $\Omega_0 h^2$, and $\Omega_b h^2$)
\begin{equation}
\sigma_8 e^{-\tau_{\rm es}} = 0.765 + 0.6(n-1) ,
\label{eq:sigma8norm}
\end{equation}
where formula (\ref{eq:sigma8norm}) is good to better than 0.5\%.
Note that at the WMAP best-fit values of $n=0.99$ and $\tau_{\rm es}=0.17$, 
this formula gives the WMAP best-fit value of $\sigma_8=0.9$.

We assume the following heuristic form for the time-dependence 
of the UV-efficiency:
\begin{equation}
\epsilon_{\rm UV} = \epsilon_{\rm UV,0} \left( 1+ A 
e^{-f_\ast/f_{\ast,\rm crit}}\right),
\end{equation}
where $f_\ast$ is the fraction of baryons in stars.  
Thus, the luminosity of each ionizing source is given by
\begin{equation}
	L(M_b) = (1-f_\ast)M_b c^2 \epsilon_{\rm esc} 
	\epsilon_{\ast} \epsilon_{\rm UV,0} t_{\rm dyn}^{-1} \left( 1+ A 
	e^{-f_\ast/f_{\ast,\rm crit}}\right) .
\end{equation}
The new parameters $A$ and $f_{\ast, \rm crit}$ define the 
time-dependence of the efficiency.  
The essential features are that at very high 
redshift, the total efficiency is $(1+A)$ times greater than at lower 
redshift, with the transition occuring at 
$f_\ast \sim f_{\ast,\rm crit}$.  
This form is motivated by 
the notion that the first stars, being metal-free, should have 
had a higher effective efficiency, but that as metals were 
produced by this generation of stars and reinjected back 
into the IGM, the efficiency would decrease.  However, we 
emphasize that this model was simply selected heuristically, 
and was not based on any detailed model of star formation 
feedback on the UV-efficiency.  
For instance, it should be noted that
the ``enhancement'' factor may be a combination of changes in
the intrinsic efficiency $\epsilon_{\rm UV}$ and the 
resolution factor $\epsilon_\ast$ related to the fraction
of cooling gas that forms stars.  
To give a sense of the meaning of $f_{\ast,\rm crit}$, 
Table~\ref{tab:ztrans} presents the 
transition redshifts $z_{\rm trans}$ 
where $f_\ast \approx (\ln 2) f_{\ast,\rm crit}$ so 
that $\epsilon_{\rm UV}(z_{\rm trans}) \approx 
\epsilon_{\rm UV,0} (1+A/2)$.  

We fix the value of $f_{\ast,\rm crit} = 10^{-4}$ (we discuss
the sensitivity to this parameter below). 
Thus, for each
point in the $n-\tau_{\rm es}$ plane, our procedure 
searching parameter space is as follows: (1) Pick value of $A$;
(2) Adjust $\epsilon_{\rm UV,0}$ so that 
$\tau_{\rm es,model}$ (calculated by model) is consistent
with $\tau_{\rm es}$;
(3) Stop if $\chi^2$ is minimized, otherwise go back to
step (1).
For each point in the $n-\tau_{\rm es}$ plane, we have a value of
a best fit $\tilde{\chi}^2(n,\tau_{\rm es})$.  We treat 
$\tilde{\chi}^2$ as approximately the $-2\ln {\cal L}$, where
${\cal L}(n,\tau_{\rm es}|f_{\ast,\rm crit})$ is the 
likelihood function, and integrate (with uniform
priors) to find find the confidence regions 
(the results are insensitive to the priors).

\begin{figure}
\plotone{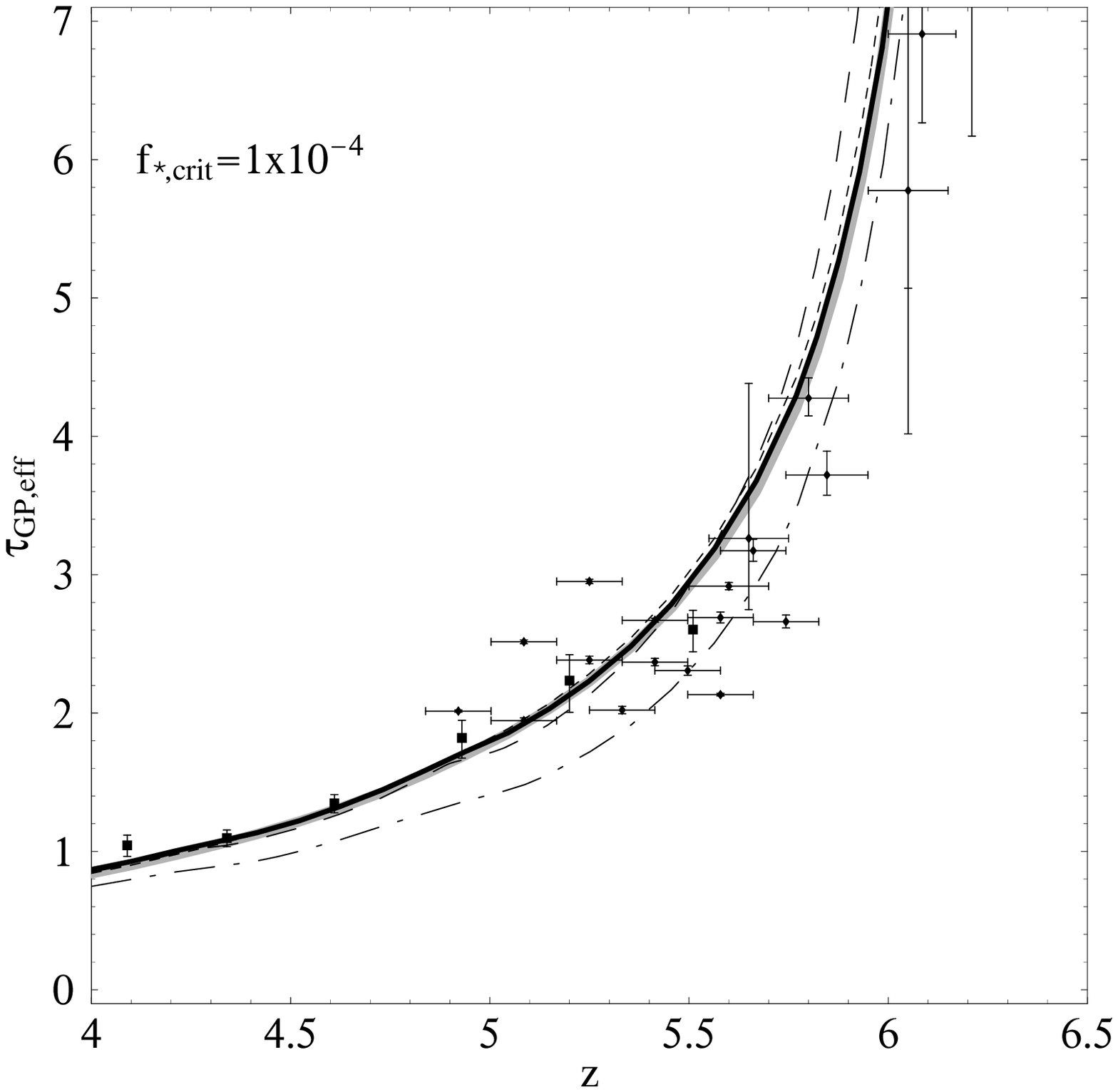}
\caption{
Plot of effective Lyman-$\alpha$ 
optical depth $\tau_{\rm eff} = -\ln({\cal T})$ 
from \citet{SC2002} compilation (boxes, error bars
are $\sigma_{\rm mean}$), 
SDSS quasars (diamonds, errors
do not include $\sigma_{\rm scatter}$), 
and model predictions for 
three model runs within the 95\% region of the $n-\tau_{\rm es}$
plane for $f_{\ast,\rm crit}=1\times 10^{-4}$, and one model
run outside of the region (i.e., ruled out by the data). 
The values of $(n,\tau_{\rm es})$, in decreasing likelihood,
are thick solid line:(0.96,0.11), short dash:(0.93,0.08), 
long dash:(0.98,0.13), dot-dash:(0.99,0.14).  For comparison, 
the thick gray line is the best-fit model for a constant
efficiency at $\Omega_0=0.27$ and $n=0.99$ ($\sigma_8=0.64$).
}
\label{fig:taueffplot}
\end{figure}

The effective optical depth for several models is shown in
Figure~\ref{fig:taueffplot}.  The three illustrative ``good-fitting''
models are in the 95\% region, and the ``badly-fitting''
model is outside the region.  Clearly, high values of
$n$ and $\tau_{\rm es}$ are ruled out by the data.

\begin{figure}
\plotone{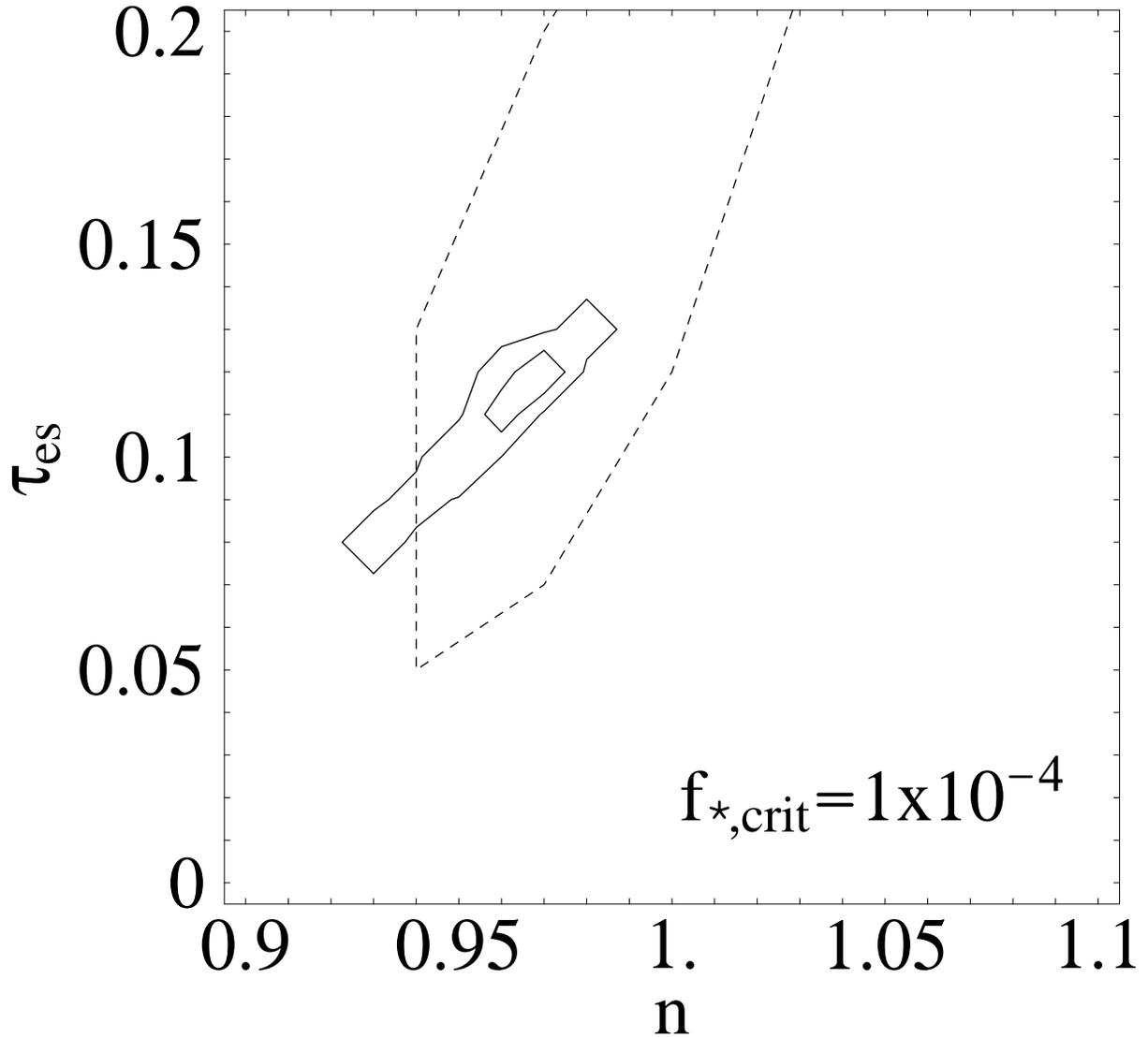}
\caption{
Constraints (68\% and 95\% contours) 
from quasar observations for a time-varying UV-efficiency 
in the $n-\tau_{\rm es}$ plane, for WMAP-normalized models with  
$f_{\ast,\rm crit} = 1\times 10^{-4}$.
Also shown (short dashed line) is approximatly the 68\% constraint
from WMAP (Spergel \etal 2003, Figure 5).
}
\label{fig:ntauresults}
\end{figure}

Considering all the constraints together, the results
are shown in Figure~\ref{fig:ntauresults}.  
Our analysis strongly favors a narrow range (95\% errors):
\begin{eqnarray}
\tau_{\rm es} & = & 0.11^{+0.02}_{-0.03}\\ 
n & = & 0.96^{+0.02}_{-0.03}
\end{eqnarray}
for $f_{\ast,\rm crit}=10^{-4}$.  
This range is consistent with the WMAP results, 
also shown in the Figure.  
The effect of changing $f_{\ast,\rm crit}$, 
is only to shift the $\tau_{\rm es}$ constraint up or
down.  For instance, the constraint for 
$f_{\ast,\rm crit}=3\times 10^{-5}$
is $\tau_{\rm es}  =  0.12^{+0.02}_{-0.03}$,
with no change in the $n$ constraint.  The results for
increasing $f_{\ast,\rm crit}$ are in the opposite direction,
decreasing the constraint on $\tau_{\rm es}$ by $\sim 0.02$ 
for a factor of 10 increase in $f_{\ast,\rm crit}$.

\section{Discussion}

\begin{figure}
\plotone{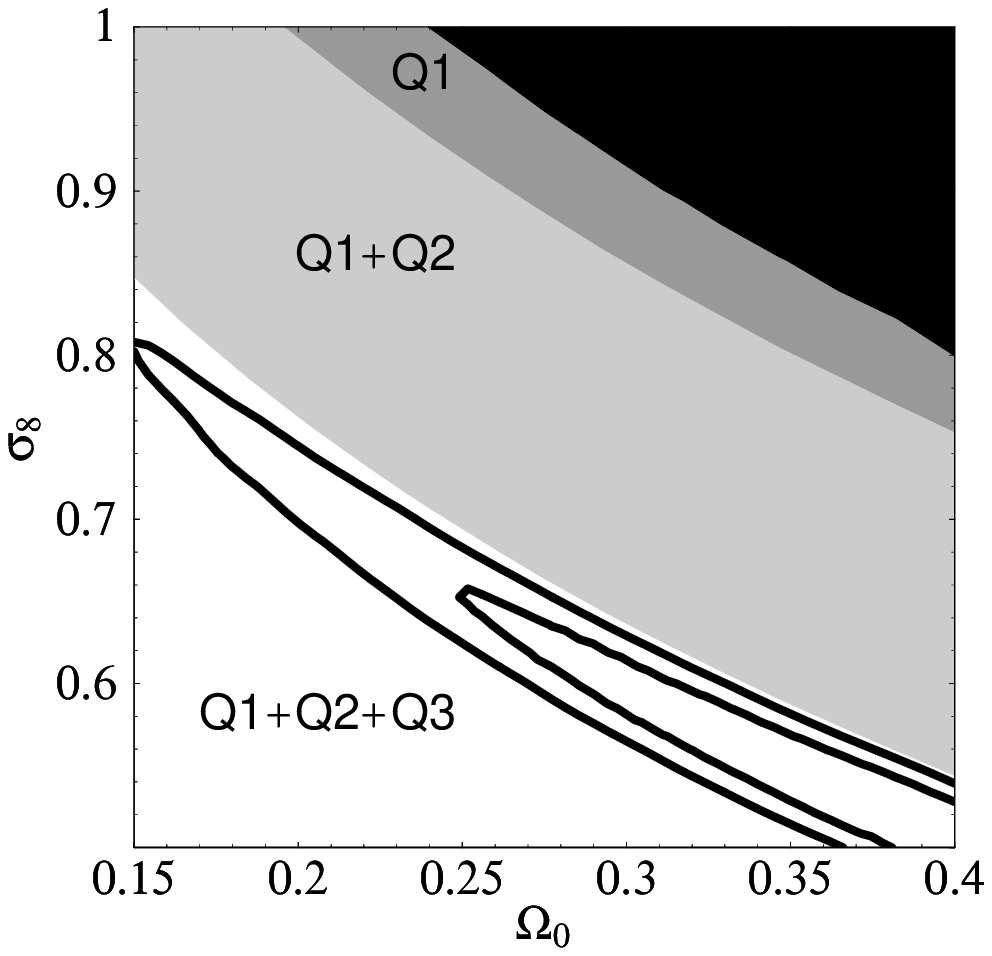}
\caption{
Summary of cosmological constraints in the $\Omega_0 - \sigma_8$ plane
for a constant UV-efficiency.  
The constraints Q1, Q2, and Q3, as described in the text, are labeled
in the shaded regions (``+'' in this context denotes ``intersection'').  
The thick solid lines are the formal 
68\% and 95\% contours using all the quasar data. }
\label{fig:mainresult}
\end{figure}

\begin{figure}
\plottwo{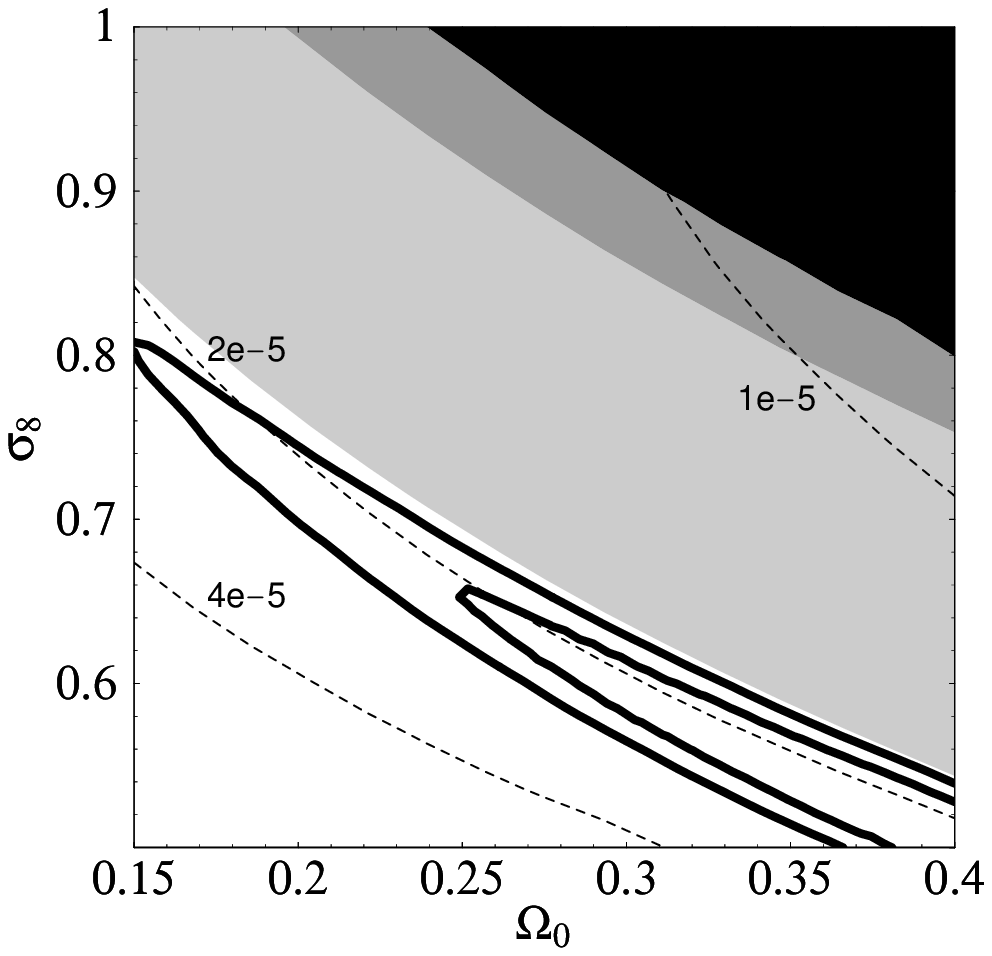}{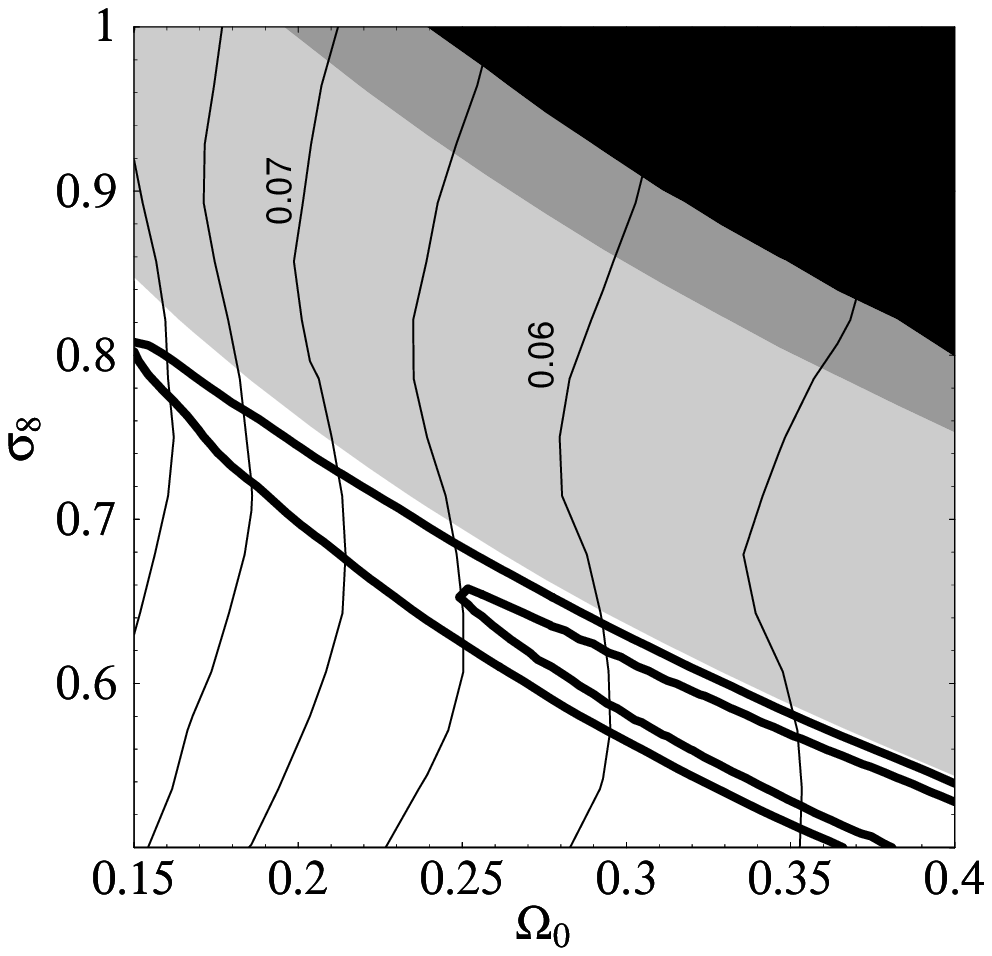}
\caption{
The values of the UV-efficiency $\epsilon_{\rm UV}$ (left panel) 
and the optical depth
to electron scattering $\tau_{\rm es}$ (right panel), overlaid on the
quasar data constraints (Figure~\ref{fig:mainresult}), for a constant
UV-efficiency.  }
\label{fig:epsandtau}
\end{figure}

In order to understand the quasar constraints, let us 
return to the constant efficiency case.  
Consider the following ``subset'' of the quasar data:
\begin{description}
\item[Q1] Requiring $z_{\rm overlap} \geq 6$ and only using the 
Ly$\alpha$ constraint from quasar J1148+5251 ($z_{\rm em} = 6.43$);
\item[Q2] Using the Ly$\beta$ constraint from
SDSS 1030+0524 ($z_{\rm em} = 6.28$);
\item[Q3] Using the \citet{SC2002} constraints at 
$z\lesssim 5$.
\end{description}
Figure~\ref{fig:mainresult} also shows the effects of combining these three 
constraints.  The intersection (Q1+Q2+Q3) subset of data appears 
to adequately depict the upper limit of the combined dataset.  
In particular, these three sets of data contain much of the 
``information'' in the full $\chi^2$ treatment.
The lower limit of the combined dataset actually also contained 
within the constraint data in Q3, but changes as a function
of $z_{\rm overlap}$. Since Q3 is the union of all these 
constraints for all values of $z_{\rm overlap}$, the lower limit
does not appear.  

This illustration shows that given the three degrees for freedom
$\Omega_0$, $\sigma_8$, and $\epsilon_{\rm UV}$, the quasar data
constrains to essentially a line in this space.  This is 
because for each $\Omega_0$, the data (Q1+Q2) at  $z\sim 6$ and 
the data (Q3) at $z\sim 4.5$ provide strong, essentially 
independent, constraints on the remaining parameters 
$\sigma_8$ and $\epsilon_{\rm UV}$.  
Of course, the ``length'' and ``width'' of the the line 
(i.e., the exact shape, as illustrated in 
Figure~\ref{fig:mainresult})
depend on the likelihood function in detail.

Now let us return to the time-varying efficiency case.  
For each value of $f_{\ast,\rm crit}$, we have four parameters 
$n$, $\tau_{\rm es}$, $\epsilon_{\rm UV,0}$, and $A$.  
Heuristically, we now have three constraints on 
the output: the data (Q1+Q2) at $z\sim 6$, the data
(Q3) at $z\sim 4.5$, and the self-consistency of 
$\tau_{\rm es}$ (value ``in'' = value ``out'').  Therefore,
once again, we expect a ``line'' in the 4-dimensional parameter
space (for fixed $f_{\ast,\rm crit}$).  As before,
the ``length'' and ``width'' of the the line will depend 
on the likelihood function in detail.  

We note also that the implied values for the normalization
in our 95\% region is $\sigma_8 \sim 0.83^{+0.03}_{-0.05}$ (based on
WMAP best-fit), which for our assumed value of $\Omega_0$ gives 
$\sigma_8\Omega_0^{0.6}=0.38^{+0.015}_{-0.025}$, just overlapping
with the cluster measurements $\sigma_8\Omega^{0.6}=0.33\pm0.03$.

The next question is what are the implications as to 
the implied values of the UV-efficiency?  Do the efficiencies
make sense?  Table~\ref{tab:parmres} shows a number of parameter
combinations from the 95\% region.  
The implied efficiency
at high redshift is on the order $\sim 10^{-4}$; a transition
occurs at $z = 15\sim20$, and the efficiency at low 
redshift is $\epsilon_{\rm UV,0} = 10^{-5.5 \sim -5}$.  
The span of these efficiencies encompasses those typically 
calculated through population synthesis of 
$10^{-5 \sim -4.5}$.  
Given uncertainties of factors of a few
in the gas collapse fraction (our resolution factor 
$\epsilon_\ast$), dust absorption (at lower redshift), etc., 
the values of the efficiencies do not seem unreasonable.  
They certainly do not approach the upper bound 
for conversion from nuclear reactions of $\sim 10^{-3}$.
\citet{Sok03} and \citet{WL03} discuss further 
the star formation efficiency 
as relating to Population II and Population III stars.

Our results are consistent with the results of \citet{Cen03}
in that we find that to reach $\tau_{\rm es} \geq 0.17$ 
requires that the spectral index is positively 
tilted with $n \gtrsim 1.02$.  In this case
the effective UV-efficiency was at least $100\times$ 
greater at $z \gg 6$.  However, our calculations indicate
that models with such high $n$ (and hence
high power spectrum normalization $\sigma_8$) are 
inconsistent with quasar transmission measurements at 
redshift $z \lesssim 5$.  

Like other authors (e.g., Haiman and Holder 2003,
Cen 2003), we find that a value of $\tau_{\rm es} = 0.17$
is inconsistent with constraints at $z\lesssim 6$ for simple 
models of reionization, and may require more exotic (though
not necessarily implausible) methods for creating
ionizing photons, such as mini-quasars or an X-ray background.  
However, we conclude that WMAP data 
taken as a whole and quasar observations at 
$z\lesssim 6$ in fact $\emph{are entirely}$ consistent for reasonable
values and time-dependence for the UV-efficiency.
Our analysis shows the importance 
of taking into account the significant degeneracy between
$\tau_{\rm es}$ and $n$ determined by WMAP.  

Having found a consistent set of evolutionary models, let us describe
their properties in slightly greater detail.  We consider the
following two models: a model with constant UV-efficiency
with $h=0.72$, $\Omega_0=0.27$, $n=0.99$, $\sigma_8=0.64$,
$\epsilon_{\rm UV}=2.2\times 10^{-5}$, and $\tau_{\rm es}=0.06)$; 
and a model
with variable efficiency with $h=0.72$, 
$\Omega_0=0.27$, $n=0.96$, $\sigma_8=0.827$,
$\epsilon_{\rm UV}=7.7\times 10^{-6}(1+17e^{-f_\ast/10^{-4}})$,
and $\tau_{\rm es}=0.11$.
These are both near peak of the likelihood distribution 
for $\Omega_0=0.27$.  

\begin{figure}
\plottwo{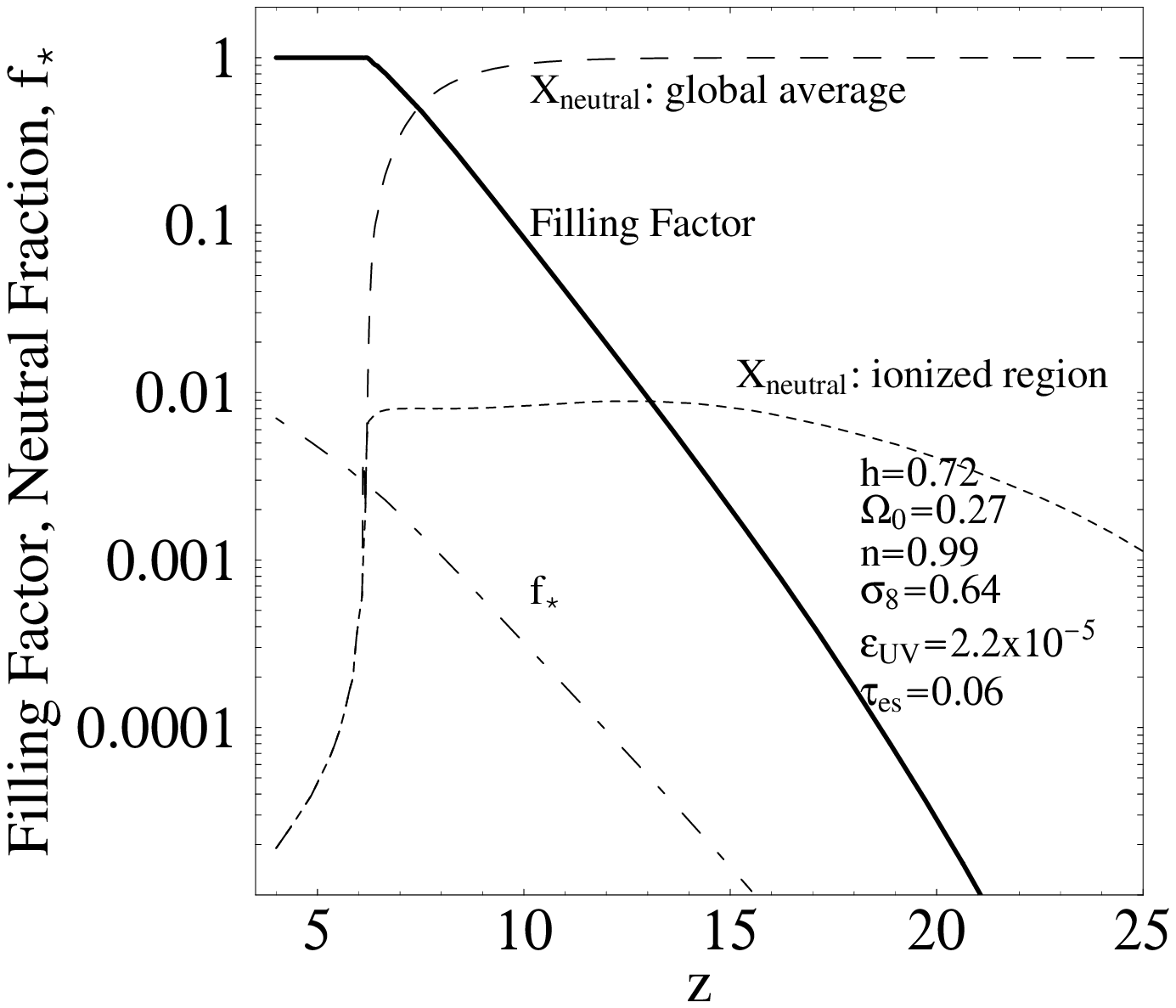}{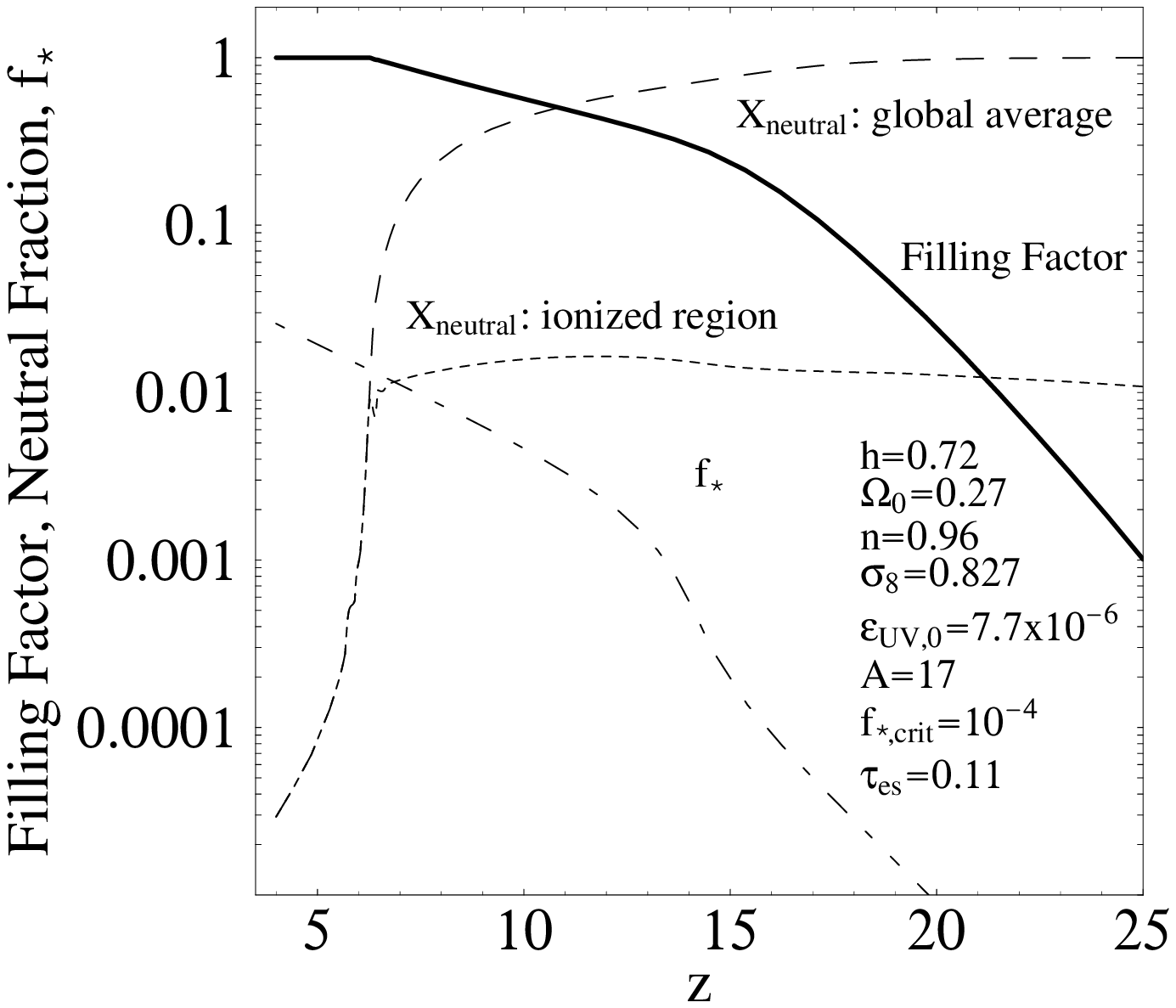}
\caption{Some basic reionization properties of the
``best fit'' models with a constant UV-efficiency (left
panel, $\tau_{\rm es}=0.06$) and with a 
time-dependent UV-efficiency (right
panel, $\tau_{\rm es}=0.11$) as a function of redshift $z$.  
Show are the filling factor (solid line), neutral hydrogen 
fraction in the ionized region (short dashed) and the 
global average neutral hydrogen faction (long dashed), and
the fraction of baryons in stars $f_\ast$ (long-short dashed).
}
\label{fig:bestfitplot}
\end{figure}

The Gunn-Peterson optical depths
are already shown in Figure~\ref{fig:taueffplot}. 
The best-fit models do not show much difference, as expected
since the parameters were fit to these data.
Figure~\ref{fig:bestfitplot} shows
the reionization properties for the two models. 
although the last phase of reionization occurs rapidly
in both models, the phase in which the average neutral
fraction drops from unity to $\sim 0.1$ takes a 
significantly longer time in the model with a 
variable UV-efficiency.  This is a ``necessary'' part
of the model in order to achieve a higher electron
scattering optical depth.  The transition from
the higher to the lower UV-efficiency is clearly
seen in the stellar baryon fraction $f_\ast$ 
at redshift $z \sim 15$.  We note that, even in the
case of $\tau_{\rm es}=0.11$, there are not two distinct
epochs of ionization, in contrast to the calculations of
\citet{Cen02a}.  Rather, there is an extended period 
during which the ionization increases from 1\% to 10\% 
before the rapid phase change at $z\sim 6$.

\begin{figure}
\plottwo{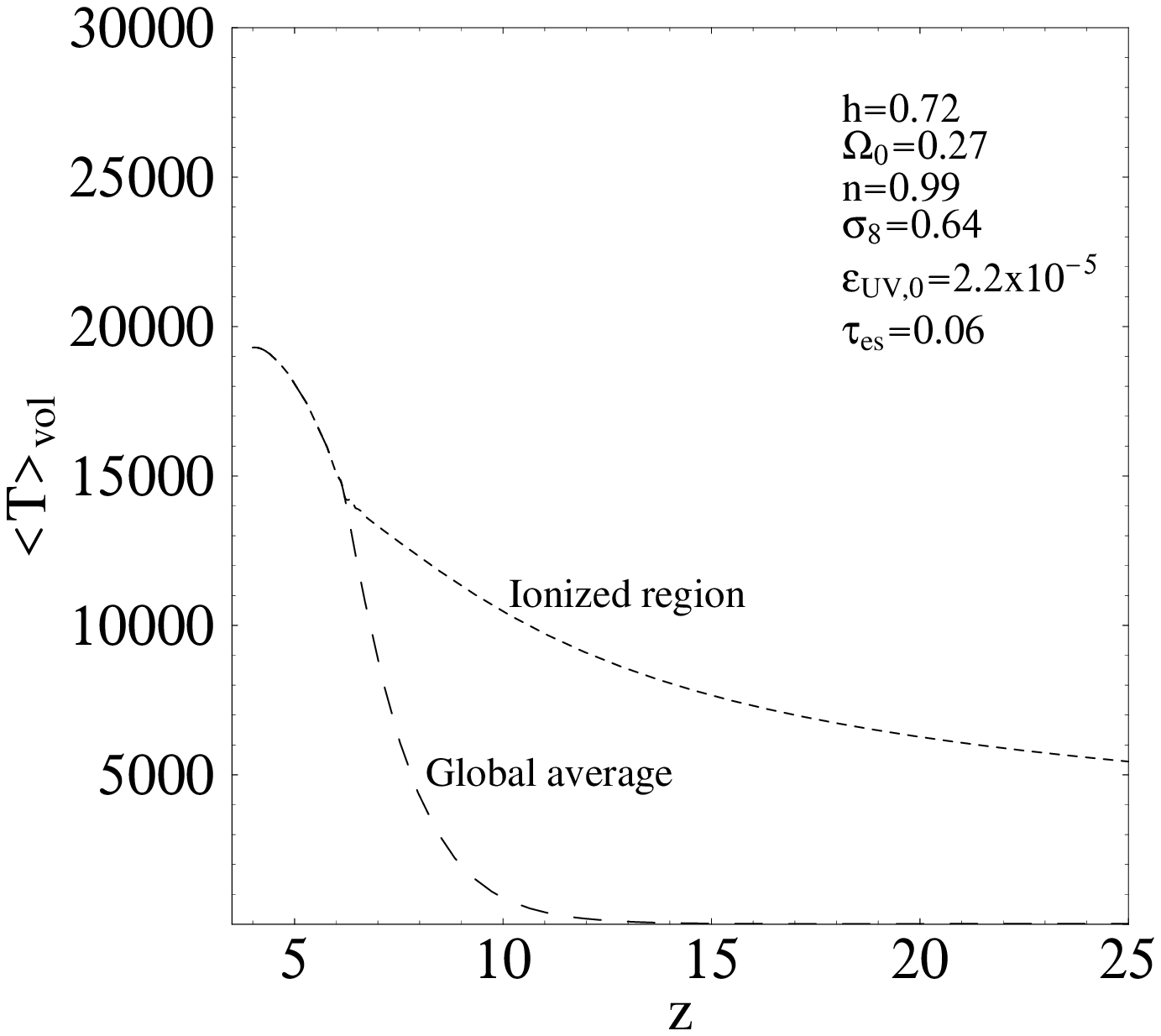}{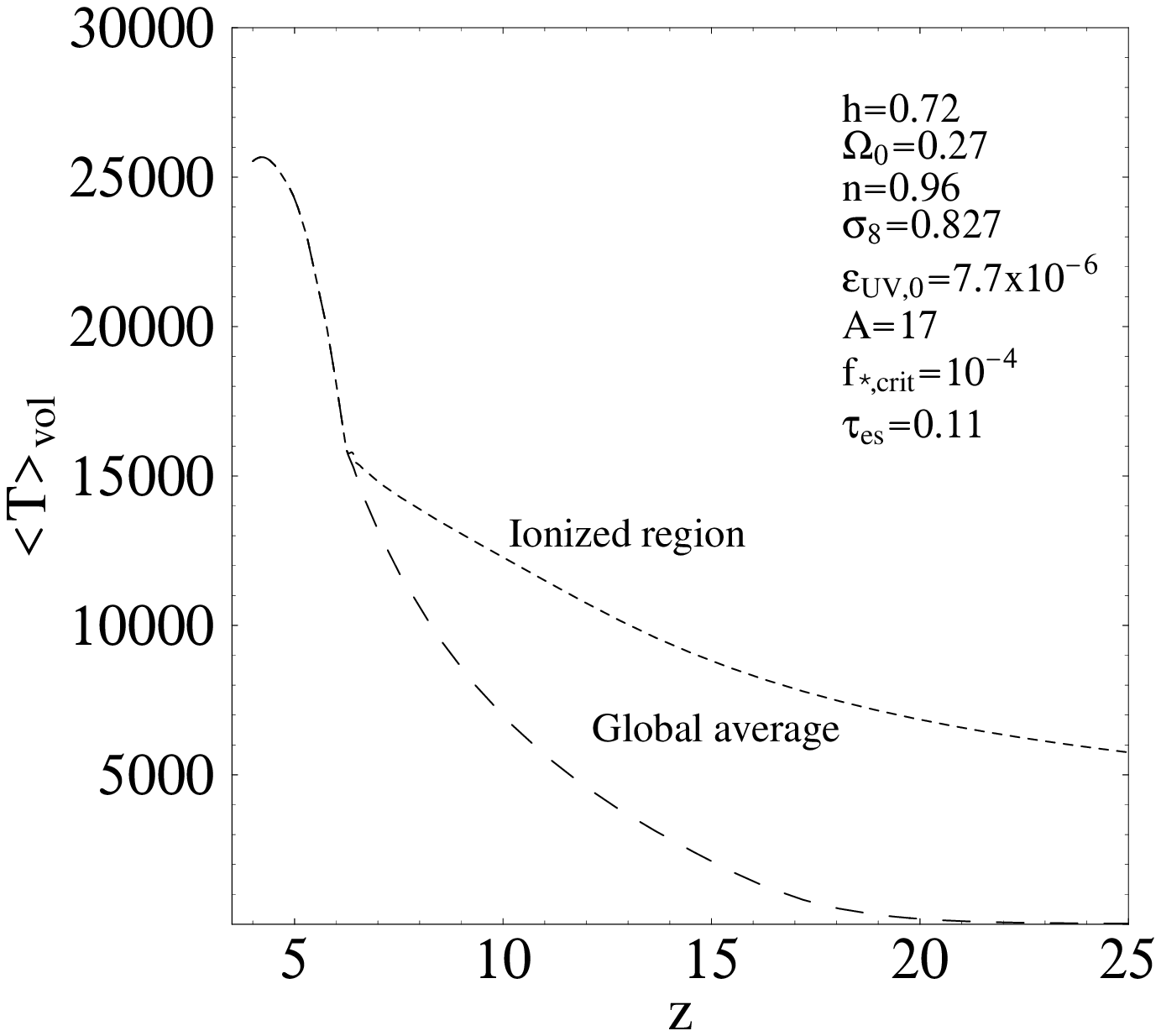}
\caption{Thermal properties of the
``best fit'' models with a constant UV-efficiency (left
panel, $\tau_{\rm es}=0.06$) and with a time-dependent UV-efficiency (right
panel, $\tau_{\rm es}=0.11$) as a function of redshift $z$.  Shown are the 
volume-average temperature in the ionized region (short dashed) and
the global volume-average (long dashed).
}
\label{fig:besttempplot}
\end{figure}

The thermal properties are show in Figure~\ref{fig:besttempplot}. 
Again, the rise in the global 
volume-averaged temperature is much more gradual
with the variable UV-efficiency.  In addition, in this
case, the final temperature is higher, but this is 
due to the higher power spectrum normalization
($\sigma_8$).  These temperatures are somewhat lower that
the ``peak'' temperatures derived by \citet{HuiH03}, but
this is probably due to the ``sudden'' reionization 
model used in those calculations.  A more gradual reionization
transition would smooth out the peak, and would appear
consistent with our calculations.  For instance, \citet{HuiH03}
consider a ``stochastic'' reheating process 
their derived temperatures at $z\sim 4$ are 
similar to ours.  

\begin{figure}
\plottwo{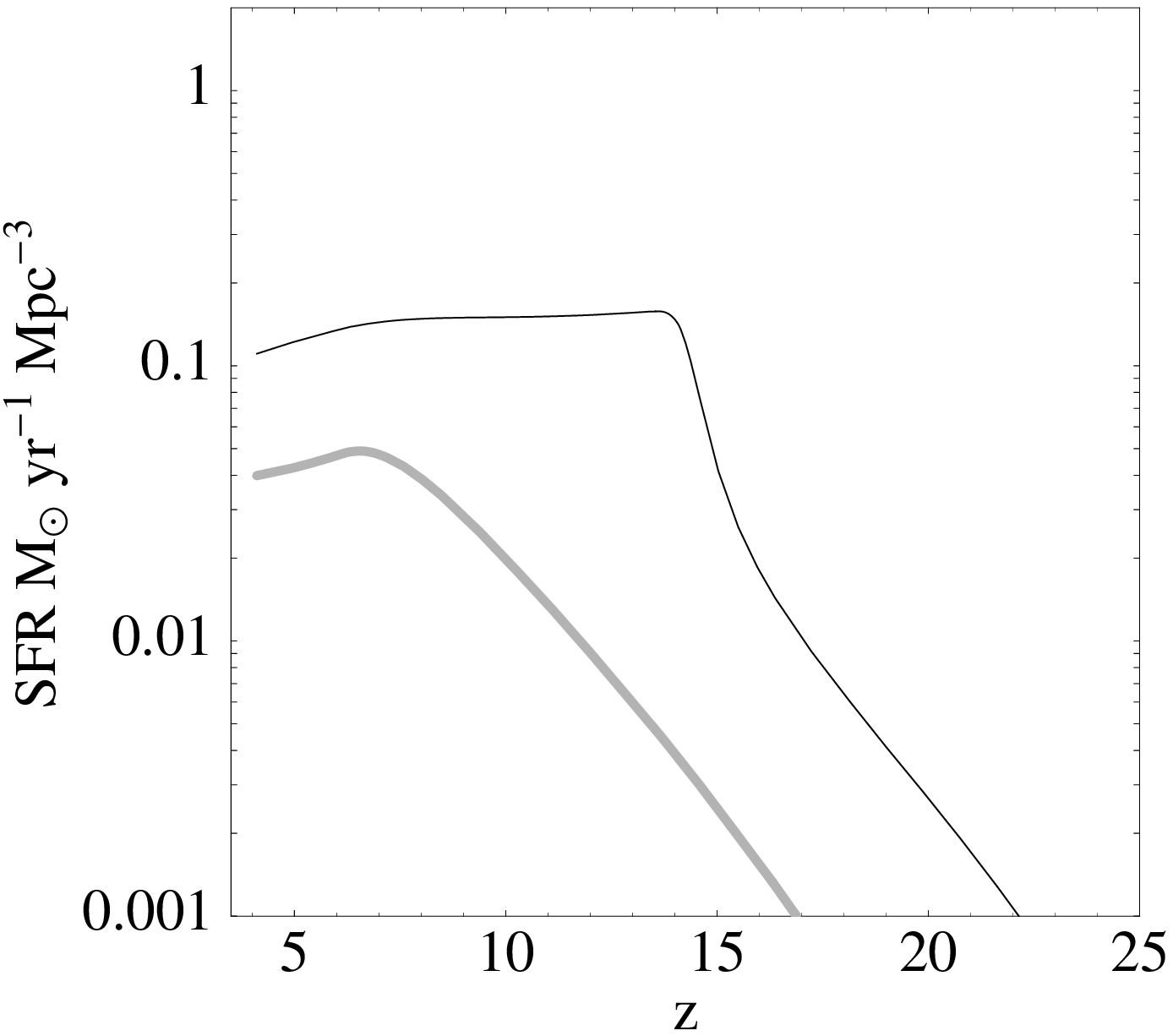}{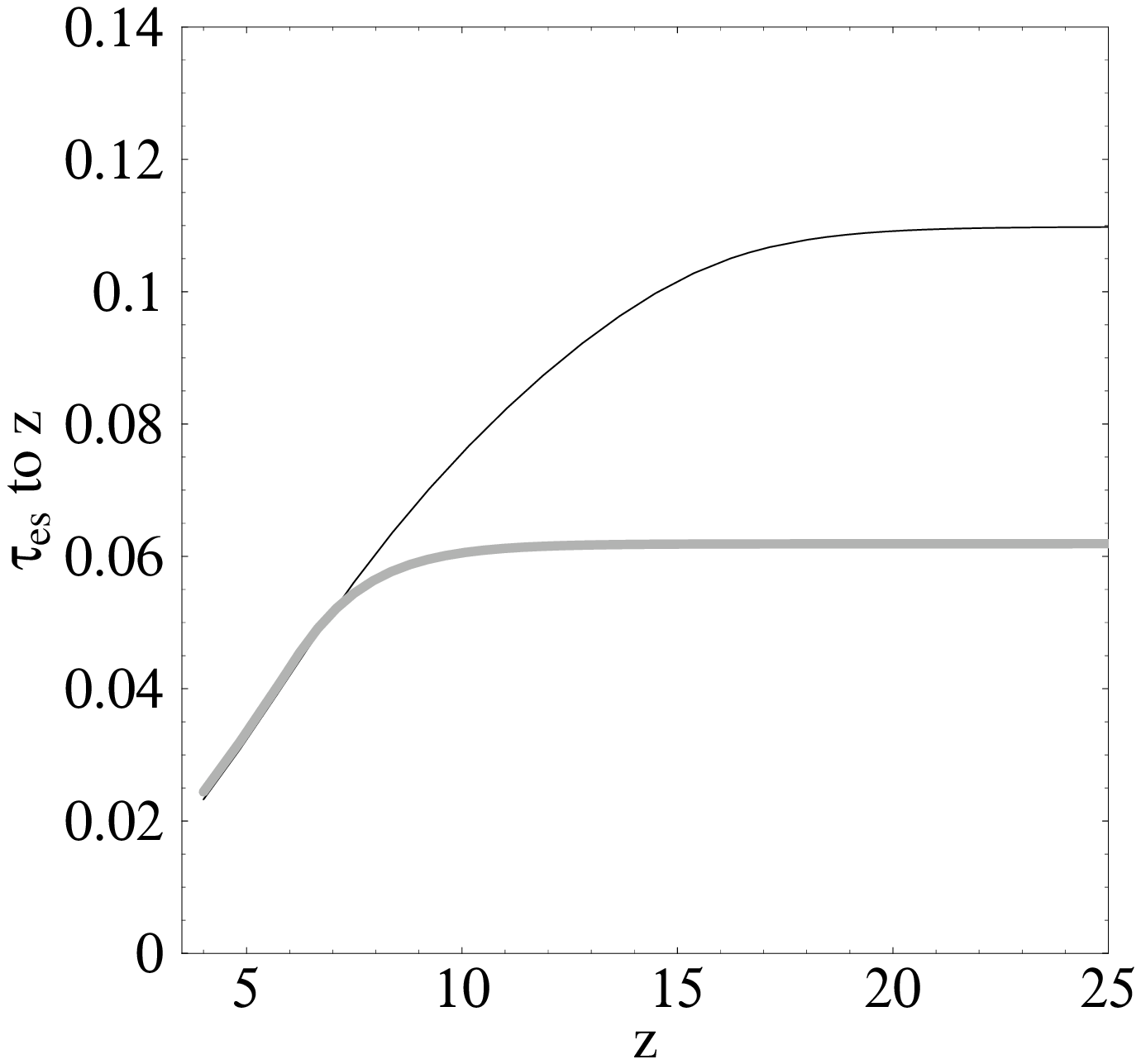}
\caption{
Star formation rate (left panel) and electron scattering
optical depth to redshift $z$ (right panel) for the
constant UV-efficiency (thick gray line, $\tau_{\rm es}=0.06$) and 
the variable UV-efficiency (thin solid line, $\tau_{\rm es}=0.11$) 
``best fit'' models.
}
\label{fig:bestsfrtauesplot}
\end{figure}

Finally, in Figure~\ref{fig:bestsfrtauesplot} we show
the star formation rate and the electron scattering optical
depth to redshift $z$.  Star formation rates at 
$z\sim 4$ from field galaxy measurements have been
reported at about $\sim 10^{-2.5 \pm 0.5}$ M$_{\sun}$ 
yr$^{-1}$ Mpc$^{-1}$ \citep{SFR99}, while a recent lower limit
of $\sim 10^{-3.1}$ M$_{\sun}$ 
yr$^{-1}$ Mpc$^{-1}$ has been reported at $z\sim 6$
\citep{SFR03}.  Our model
results  imply that there is exists 
a large population of unobserved sources at $z\sim 6$.  
We not that quasar observations alone require that $\tau_{\rm es}$
has reached value of $\sim 0.05$ at $z\sim 7$.
Finally, the electron
scattering optical depth shows that in the constant
efficiency case, the full optical depth is reached by 
$z\sim 10$, while in the variable efficiency case, the
full optical depth is not reached until $z\sim 20$.
This is an observational signature that could be detected
by future CMB experiments.

\section{Summary and Conclusions}

We derive constraints on several cosmological parameters based on
observations of Gunn-Peterson absorption in high-redshift quasars
and WMAP observations \citep{WMAP}.
We use a semi-analytic model for reionization \citep{Chiu00}
that takes into account 
a number of important physical processes both within collapsing halos 
(e.g., H$_2$ cooling) and in the intergalactic medium (e.g., 
H$_2$ cooling, Compton cooling, and photoionization heating).  
The model is also calibrated to hydrodynamic simulations.  
We also develop a method
for estimating the gas PDF, which is important for properly 
calculating the mean absorption in the IGM, 
as a function of cosmological parameters. 

We find that the Gunn-Peterson absorption
data provide constraints on the power spectrum at small
scales in a manner similar to that derived from the cluster mass function.
Assuming that the efficiency of producing UV photons per baryon is 
constant, the constraint takes on the form 
$\sigma_8\Omega_0^{0.5} = 0.33 \pm 0.01$  
assuming a flat, $\Lambda$-dominated universe with
$h=0.72$, $n=0.99$, and $\Omega_b h^2 = 0.024$. 
The best
fit for the WMAP data (marginalized over all parameters) is reported
as $\sigma_8\Omega_0^{0.5}=0.48\pm0.12$, which differs by slightly
more than $1-\sigma$. However, the derived value for the optical
depth to last-scattering $\tau_{\rm es} \approx 0.06$ differs
signficantly from the WMAP-determined value of 
$\tau_{\rm es} = 0.17\pm 0.04$.  

Since the WMAP constraints on $\tau_{\rm es}$ are somewhat degenerate
with the value of the spectral index $n$ \citep{Setal03}, 
we then let the primordial spectral index $n$ float while 
fixing the best fit WMAP determination of $\Omega_0 h^2 = 0.14$ 
and normalizing the power spectrum using WMAP.
In addition, we allow the UV-efficiency to have time-dependence.
Combining the WMAP constraints with the quasar transmission data,
our analysis favors then a model with 
$\tau_{\rm es}=0.11^{+0.02}_{-0.03}$,
$n = 0.96^{+0.02}_{-0.03}$, implying $\sigma_8=0.83^{+0.03}_{-0.05}$
(all at 95\% confidence), and an effective UV-efficiency 
that was at least $\sim 10\times$ greater at $z \gg 6$.  
If future observations confirm this range for the optical depth
to electron scattering $\tau_{\rm es}$, then it would appear that
no more ``exotic'' sources of UV-photons are necessary.  We are unable
to find a model that is consistent with all observational and
physical constraints that has an electron scattering optical 
depth within the 1-$\sigma$ range given by the WMAP team.
If the EE data finally require a value $\tau_{\rm es} \gtrsim 0.17$,
then more exotic sources of early ionizing photons will be required
that those considered in this paper.

\acknowledgments

The authors would like to thank Renyue Cen 
kindly providing his simulation outputs, and Martin Rees,
Massimo Ricotti, and David Spergel for useful discussions.   
XF thanks support from the University of Arizona and
a Sloan Research Fellowship



\appendix
\section{Approximate Model for Escape Fraction}

In this appendix, we consider a simple Str\"omgren sphere approximation
for the escape fraction.  The basic conditions for star forming 
halos in our semi-analytic model are as follows:
\begin{enumerate}
\item In halos greater than the Jeans mass, the baryonic mass 
will collapse to a singular isothermal sphere with 
$\rho_b = \frac{1}{3}\Delta_v \bar{\rho}_b (r_v/r)^2$, where 
$\Delta_v \sim 178$ is the virial overdensity, $\bar{\rho}_b$ 
is the mean baryonic density, and $r_v$ is the virial radius.
Here the total baryonic mass is 
$M_b = \frac{4\pi}{3}\Delta_v \bar{\rho}_b r_v^3$. 
\item In halos where the cooling time $t_{c}$ is less than the
dynamical time $t_{d}$, there will be star formation.
\item The rate of UV photon production per unit baryonic mass 
is $\theta = c^2 \epsilon_\ast \epsilon_{\rm UV} E_0^{-1} t_d^{-1}$,
where $\epsilon_{\rm UV}$ is the mass to UV energy efficiency and
$E_0 = 13.6$~eV.  Here $\epsilon_\ast$ is the fraction of cooling
baryons in the halo that are forming stars (the ``resolution'' factor),
and $\epsilon_{\rm UV}$ is the fundamental conversion efficiency
from stellar baryons to UV photons.
\end{enumerate}

In regions that are optically thick, so that all 
photons are absorbed locally, then 
local ionization equilibrium would imply that 
$\theta \rho_b = n_{\rm H}^2 x^2 \alpha$, where $n_{\rm H}$ 
is the total hydrogen density, $x$ is the ionization fraction, 
and $\alpha$ is the recombination coefficient.  This implies 
that the ionization fraction is 
$x = \sqrt{\theta\rho_b/\alpha}/n_{\rm H} \propto r$. 
Therefore, as $r\rightarrow 0$, the ionized fraction 
also tends to 0.  

To check for self consistency, consider the optical depth
over a central region.  The average 
density over the sphere $r$ is 
$\langle \rho_b(r)\rangle = \Delta_v \bar{\rho}_b (r_v/r)^2$. 
Optical depth over this region is approximately 
$\langle \tau \rangle \sim \langle n_{\rm H} (1-x)\rangle \sigma r$,
where $\sigma$ is an effective cross section.  Since 
$\langle n_{\rm H} \rangle \sim r^{-2}$, 
this implies that $\langle \tau \rangle \sim (1-x) r^{-1}$.  
Therefore, as long as $(1-x)$ does not vanish,
then the optical depth will become very large as $r\rightarrow 0$.
Above, we established that $x \rightarrow 0$ as $r\rightarrow 0$ 
for an optically thick region.  We therefore have a self consistent
picture for the central part of the halo.

Now what is the condition so that the entire halo is optically
thick?  For a completely neutral halo, 
$\tau \sim \Delta_v \bar{\rho}_b \sigma r_v / m_{\rm eff}$.  Plugging
in numbers gives $\tau \sim r_v (1+z)^3 / 10$~kpc (in physical units).  
In comoving units, $\tau \propto (1+z)^2$.  Therefore at any redshift 
before full reionization ($z \gtrsim 6$), all halos of concern are 
optically thick if fully neutral.  They only become optically thin
if the neutral fraction is small.

Now consider a model based on a central source 
approximation.  Let $S(r)$ be the number of ionizing 
photons emitted by a central source which pass through a
sphere of radius $r$.  The standard 
equation for $S(r)$ is 
given by 
\begin{equation}
\frac{\partial S(r)}{\partial r} = -4\pi r^2 n_{\rm H}^2 x^2 \alpha .
\end{equation}
Using the definitions above, we obtain
\begin{equation}
\frac{\partial S(r)}{\partial r} = -4\pi   
\left(\frac{\Delta_v\bar{\rho}_b r_v^2}{3 m_{\rm eff}}\right)^2 
\frac{x^2 \alpha}{r^2}  .
\end{equation}
Now consider, the number of photons produced within $r$,
${\cal S}(r) = \theta M(<r) $.  Differentiating this with 
respect to $r$ gives
\begin{equation}
\frac{\partial {\cal S}(r)}{\partial r} = 
\frac{4\pi}{3} \theta \Delta_v \bar{\rho}_b r_v^2 .
\end{equation}
Now let us make the ansatz that all these photons can be considered
to be radially emitted.  Then these two equations can be \emph{combined}
to yield 
\begin{equation}
\frac{\partial S(r)}{\partial r} = 
\frac{4\pi}{3} \Delta_v \bar{\rho}_b r_v^2 
\left( 
1 -   
\frac{\Delta_v\bar{\rho}_b r_v^2}{3 \theta m_{\rm eff}^2} 
\frac{x^2 \alpha}{r^2}  \right)
\label{eq:escapefracdiffeq}
\end{equation}
Because we know that the halo is optically thick if $x < 1$, 
let us consider the following prescription: as long as $x < 1$, 
all photons are absorbed locally, so $\frac{\partial S}{\partial r} = 0$.
This leads to the expression for the ionized fraction as a function 
of radius
\begin{equation}
x(r) = {\rm Min}\left[\frac{r}{r_v}
	\sqrt{\frac{3 \theta m_{\rm eff}^2}{\Delta_v\bar{\rho}_b\alpha}} 
	,1\right].
\end{equation}
If we define quenching fraction
\begin{equation}
\eta \equiv \sqrt{\frac{\Delta_v\bar{\rho}_b\alpha}{3\theta m_{\rm eff}^2}},
\end{equation}
then $\eta \geq 1$ means that all of the photons are absorbed within
the halo.  If $\eta < 1$, then the ionized fraction has reached its
maximum value of 1 at the ``quenching radius'' $r_q = \eta r_v$. 
Note importantly that $\eta$ depends only on redshift through 
the background density $\bar{\rho}_b$ and the dynamical time 
in $\theta$, and is essentially 
\emph{the same for all halos}.  (Actually, 
there is a weak dependence on the halo temperature through the
recombination rate $\alpha$.)

If $\eta < 1$, then we use equation (\ref{eq:escapefracdiffeq}) 
with $x=1$ and the boundary condition
$S(r_q) = 0$ to solve for $S(r)$ from $r_q$ to $r_v$.  The result is 
\begin{equation}
S(r_v) = \frac{4\pi}{3} \theta \Delta_v \bar{\rho}_b r_v^3 (1-\eta)^2
	= M_b \theta  (1-\eta)^2.
\end{equation}
Thus, the factor $(1-\eta)^2$ is the fraction of UV photons created
in the halo that actually escape --- the escape fraction 
$\epsilon_{\rm esc}$.  
At high redshift, when $\eta \geq 1$, 
this factor is 0 --- all the photons are consumed within the halo. 

To illustrate, consider the ``best fit'' model with time-variable
UV-efficiency described
in the main text.  Using a temperature of $10^4$ Kelvin, $\Omega_0h^2=0.14$, 
$\Omega_bh^2=0.024$, the quenching fraction is
\begin{equation}
\eta \sim \frac{10^{-4} (1+z)^{0.75}}{\sqrt{\epsilon_\ast \epsilon_{\rm UV}}}.
\end{equation}
In our ``best fit'' model, we fixed $\epsilon_\ast = 0.03$, while
$\epsilon_{\rm UV}$ ranged from $\sim 0.8\times 10^{-5}$ 
at low redshift to $1.3\times 10^{-4}$ at high redshift.
At very high redshift, we have
\begin{equation}
\eta \approx \left(\frac{1+z}{57}\right)^{0.75}.
\end{equation}
This implies that at very high redshift, when $\eta \geq 1$, 
all the photons are absorbed locally so that the escape fraction
is essentially 0.  
This is to expected since densities are much higher at high
redshift.  At redshifts near transition $\sim 15$, when
$\epsilon_{\rm UV} \sim 6\times 10^{-5}$, we have
\begin{equation}
\eta \approx \left(\frac{1+z}{37}\right)^{0.75},
\end{equation}
implying an escape fraction $\sim 0.2$, comparable to
values used in other semianalytic models.  
At $z\sim 6$, we have 
\begin{equation}
\eta \approx \left(\frac{1+z}{9}\right)^{0.75},
\end{equation}
implying an escape fraction $\epsilon_{\rm esc} \sim 0.03$.
These values are all within the range used by others in
semi-analytic models (e.g., Cen 2003; Wiyithe and Loeb 2002; 
Haiman and Holder 2003).




\clearpage


\clearpage

\begin{deluxetable}{ccccc}
\tabletypesize{\scriptsize}
\tablecaption{Fit Parameters for Gas PDF Derived from Simulations 
\label{tab:gaspdfparms}}
\tablewidth{0pt}
\tablehead{
\colhead{Redshift} & \multicolumn{2}{c}{MHR\tablenotemark{a}} & 
	\multicolumn{2}{c}{Cen\tablenotemark{b}} \\
\cline{2-5}
\colhead{} & \colhead{$\beta$} & \colhead{$\delta_0$} &
\colhead{$\beta$} & \colhead{$\delta_0$}}
\startdata
6 & 2.50 & 1.09 & 2.52 & 2.6 \\
5 & ---  & ---  & 2.41 & 3.0 \\
4 & 2.48 & 1.53 & 2.25 & 3.6 \\
3 & 2.35 & 1.89 & ---  & --- \\
2 & 2.23 & 2.54 & ---  & --- 
\enddata
\tablenotetext{a}{Miralda-Escud\'e \etal\ 2000}
\tablenotetext{b}{Cen 2002b}
\end{deluxetable}

\begin{deluxetable}{ccccl}
\tabletypesize{\scriptsize}
\tablecaption{
Compilation of Transmission Data from \citet{SC2002}
\label{tab:transmissiondatacowie}}
\tablewidth{0pt}
\tablehead{
\colhead{Redshift} & \colhead{Mean Transmission ${\cal T}$} & 
\colhead{$\sigma_{\rm scatter}$} &
\colhead{$\sigma_{\rm mean} = \sigma_{\rm scatter}/\sqrt{N}$} &
\colhead{$N$}
}
\startdata
$ 4.09 $ & $0.352$ & $0.352$ & $0.027$ & 15 \\
$ 4.34 $ & $0.334$ & $0.334$ & $0.020$ & 20 \\
$ 4.61 $ & $0.260$ & $0.260$ & $0.017$ & 15 \\
$ 4.93 $ & $0.162$ & $0.162$ & $0.022$ & 5 \\
$ 5.20 $ & $0.107$ & $0.107$ & $0.022$ & 8 \\
$ 5.51 $ & $0.074$ & $0.074$ & $0.011$ & 7 
\enddata
\end{deluxetable}

\begin{deluxetable}{ccclll}
\tabletypesize{\scriptsize}
\tablecaption{
Transmission Data
\label{tab:transmissiondata}}
\tablewidth{0pt}
\tablehead{
\colhead{Redshift Range} 
& \colhead{Transmission $T \pm \sigma_{\rm meas}$} & 
\colhead{$\sigma_{\rm scatter}$} &
\colhead{Reference} &
\colhead{SDSS quasar} &
\colhead{Comments} }
\startdata
$ 4.84 - 5.00 $ & $0.1334 \pm 0.0011$ & $0.046$ & \tablenotemark{a} 
	& J0836+0054 & \\
$ 5.00 - 5.17 $ & $0.0809 \pm 0.0011$ & $0.055$ & \tablenotemark{a} 
	& J0836+0054 & \\
$ 5.17 - 5.33 $ & $0.0523 \pm 0.0008$ & $0.055$ &  \tablenotemark{a} 
	& J0836+0054 & \\
$ 5.33 - 5.50 $ & $0.0692 \pm 0.0010$ & $0.029$ &  \tablenotemark{a} 
	& J0836+0054 & \\
$ 5.50 - 5.66 $ & $0.1185 \pm 0.0011$ & $0.029$ &  \tablenotemark{a} 
	& J0836+0054 & \\
$ 5.25 - 5.41 $ & $0.1324 \pm 0.0036$ & $0.029$ &  \tablenotemark{a} 
	& J1030+0524 & \\
$ 5.41 - 5.58 $ & $0.0996 \pm 0.0033$ & $0.029$ &  \tablenotemark{a} 
	& J1030+0524 & \\
$ 5.58 - 5.74 $ & $0.0418 \pm 0.0033$ & $0.020$ &  \tablenotemark{a} 
	& J1030+0524 & \\
$ 5.74 - 5.95 $ & $0.0242 \pm 0.0038$ & $0.020$ &  \tablenotemark{a} 
	& J1030+0524 & \\
$ 6.0 - 6.17 $ & $0.0010 \pm 0.0009$ & $0.0005$ & \tablenotemark{b} 
	& J1030+0524 & \\
$ 6.0 - 6.17$ & $0.0043 \pm 0.0088$ & --- & \tablenotemark{b} 
	& J1030+0524 & Lyman-$\beta$ \\
$ 5.95 - 6.15 $ & $0.0031 \pm 0.0149$ & $0.0005$ & \tablenotemark{c} 
	& J1048+4637 & \\
$ 5.50 - 5.70 $ & $0.0541 \pm 0.0014$ & $0.029$ & \tablenotemark{a} 
	& J1148+5251 & \\
$ 5.70 - 5.90 $ & $0.0139 \pm 0.0019$ & $0.020$ & \tablenotemark{a} 
	& J1148+5251 & \\
$ 6.0 - 6.10 $ & $0.0 \pm 0.0063$ & --- & \tablenotemark{b} 
	& J1148+5251 & \\
$ 6.0 - 6.10 $ & $0.0 \pm 0.335$ & --- & \tablenotemark{b} 
	& J1148+5251 & Lyman-$\beta$\\
$ 6.10 - 6.32 $ & $0.0 \pm 0.0021$ & --- & \tablenotemark{b} 
	& J1148+5251 & \\
$ 6.10 - 6.32 $ & $0.0 \pm 0.051$ & --- & \tablenotemark{b} 
	& J1148+5251 & Lyman-$\beta$\\
$ 5.00 - 5.17 $ & $0.1429 \pm 0.0027$ & $0.055$ & \tablenotemark{a} 
	& J1306+0356 & \\
$ 5.17 - 5.33 $ & $0.0922 \pm 0.0025$ & $0.055$ & \tablenotemark{a} 
	& J1306+0356 & \\
$ 5.33 - 5.50 $ & $0.0936 \pm 0.0025$ & $0.029$ & \tablenotemark{a} 
	& J1306+0356 & \\
$ 5.50 - 5.66 $ & $0.0679 \pm 0.0027$ & $0.029$ & \tablenotemark{a} 
	& J1306+0356 & \\
$ 5.66 - 5.83 $ & $0.0699 \pm 0.0033$ & $0.020$ & \tablenotemark{a} 
	& J1306+0356 & \\
$ 5.55 - 5.75 $ & $0.0383 \pm 0.0258$ & $0.020$ & \tablenotemark{c} 
	& J1630+4027 & 
\enddata
\tablecomments{Lyman-$\alpha$ unless otherwise noted}
\tablenotetext{a}{Becker \etal\ (2001) }
\tablenotetext{b}{White \etal\ (2003).  For J1148+5251, these are
conservative upper limits based on the hypothesis of
Ly-$\alpha$ emission from an intervening galaxy. }
\tablenotetext{c}{Fan \etal\ (2003) }
\end{deluxetable}

\begin{deluxetable}{cccccc}
\tabletypesize{\scriptsize}
\tablecaption{$z_{\rm trans}$ as a Function of $n$ and $f_{\ast,\rm crit}$
for Semi-analytic Runs
\label{tab:ztrans}}
\tablewidth{0pt}
\tablehead{
\colhead{$f_{\ast,\rm crit}$} & 
	\multicolumn{5}{c}{$z_{\rm trans}$ for $n=$} \\
\cline{2-6} 
\colhead{} & \colhead{0.94} & \colhead{0.97} &
\colhead{1.00} & \colhead{1.03} & \colhead{1.06}}
\startdata
$3\times 10^{-5}$ & 16.5 & 20 & 24.5 & 30 & 35 \\
$1\times 10^{-4}$ & 14.5 & 18 & 22   & 27 & 32 \\
$3\times 10^{-4}$ & 12.5 & 15.5 & 19 & 23.5 & 28 \\
$1\times 10^{-3}$ & 10.5 & 13 & 16 & 20 & 24 
\enddata
\end{deluxetable}

\begin{deluxetable}{ccccc}
\tabletypesize{\scriptsize}
\tablecaption{Parameter Combinations in the 95\% Confidence Region
\label{tab:parmres}}
\tablewidth{0pt}
\tablehead{
\colhead{$\tau_{\rm es}$} & 
	\multicolumn{4}{c}{$f_{\ast,\rm crit} = 10^{-4}$} \\
\cline{2-5} 
\colhead{} & \colhead{$n$} & \colhead{$\epsilon_{\rm UV,0} (\times 10^{-4})$} &
\colhead{$A$} & \colhead{$\epsilon_{{\rm UV,high}\ z} (\times 10^{-4})$ }}
\startdata
$0.09$ & 0.94 & 0.12 & 7.8 & 0.92  \\
$0.10$ & 0.95 & 0.096 & 12  & 1.1  \\
$0.11$ & 0.96 & 0.077 & 17 & 1.3  \\
$0.12$ & 0.97 & 0.058 & 26 & 1.5  
\enddata
\end{deluxetable}




\end{document}